\documentclass[11pt, a4paper, logo, copyright, nonumbering]{lti}
\usepackage[authoryear, sort&compress, round]{natbib}
\usepackage{dblfloatfix}
\usepackage{ulem}
\usepackage{caption}
\usepackage{dramatist}
\usepackage{xspace}
\usepackage{pifont} 
\usepackage{multirow}
\usepackage{tcolorbox}
\usepackage{xltabular}
\usepackage{longtable}
\usepackage{hyperref}
\usepackage{tikz}
\usetikzlibrary{positioning}
\usepackage{amsfonts}
\usepackage{amsmath}
\usepackage{amssymb}
\usepackage{lineno}
\usepackage{multirow}
\usepackage{adjustbox}

\usepackage[bottom]{footmisc}

\usepackage{CJKutf8}
\usepackage{graphicx}
\usepackage{subcaption}
\usepackage{setspace}

\usepackage{dsfont}
\usepackage{array}
\usepackage{tabularx}
\usepackage{xcolor}
\usepackage{wrapfig}
\usepackage{booktabs}

\definecolor{ltired}{HTML}{780000}

\usepackage{lipsum}
\usepackage{listings}
\usepackage{fvextra}
\usepackage{multicol}
\usepackage{makecell}
\usepackage{forest}
\usepackage{xcolor}
\usepackage{dirtree}

\makeatletter
\def\@BTrule[#1]{%
  \ifx\longtable\undefined
    \let\@BTswitch\@BTnormal
  \else\ifx\hline\LT@hline
    \nobreak
    \let\@BTswitch\@BLTrule
  \else
     \let\@BTswitch\@BTnormal
  \fi\fi
  \global\@thisrulewidth=#1\relax
  \ifnum\@thisruleclass=\tw@\vskip\@aboverulesep\else
  \ifnum\@lastruleclass=\z@\vskip\@aboverulesep\else
  \ifnum\@lastruleclass=\@ne\vskip\doublerulesep\fi\fi\fi
  \@BTswitch}
\makeatother

\addto\extrasenglish{
}

 {\begin{list}{}%
         {\setlength{\leftmargin}{#1}}%
         \item[]%
 }
 {\end{list}}
 
\usepackage{color-edits}
\addauthor{vc}{blue}
\addauthor{rs}{brown}
\addauthor{am}{red}
\addauthor{wc}{purple}
\addauthor{bv}{magenta}
\addauthor{sw}{orange}
\addauthor{wc}{green}

\title{\centering Comparing Developer and LLM Biases in Code Evaluation}

\author{\centering
\normalsize{\textbf{Aditya Mittal$^{*}$ \quad  Ryan Shar$^{*}$ \quad Zichu Wu \quad  Shyam Agarwal \\Tongshuang Wu \quad  Chris Donahue \quad  Ameet Talwalkar \quad  \\Wayne Chi${^\dagger}$ \quad  Valerie Chen${^\dagger}$}} 
\\
\vspace{0.5em}
Carnegie Mellon University
\\
\vspace{0.5em}
\texttt{\{adityamittal307, ryan.shar01\}@gmail.com}
}

\renewcommand{\phi}{\varphi}

\renewcommand{\epsilon}{\varepsilon}
\renewcommand{\imath}{\mathrm{i}}

\newlength{\restsubwidth}
\newlength{\restsubheight}
\newlength{\restsubmoreheight}
\newcommand{\rest}[2]{%
        \settowidth{\restsubwidth}{\ensuremath{#2}}
        \settoheight{\restsubheight}{\ensuremath{{}_{#2}}}
        \ensuremath{{#1\hskip 0.5pt}_{\vrule\kern2pt\parbox[b][%
        4pt][b]{\the\restsubwidth}{%
                        \ensuremath{{}_{#2}}}}}
        }

\begin{abstract}
LLMs are increasingly used as judges in code applications, yet their efficacy remains untested in realistic code settings that often feature partial context and ambiguous intent.
We present \textbf{TRACE} (\textbf{T}ool for \textbf{R}ubric \textbf{A}nalysis in \textbf{C}ode \textbf{E}valuation), a framework that evaluates LLM judges’ ability to predict human preferences and automatically extracts rubric items to highlight systematic differences in how humans and models weight these criteria.
Across three modalities---IDE autocompletion, chat-based programming, and instructed code editing---we benchmark 13 judge models spanning general-purpose LLMs, specialized judge models, and reward models. Among these models, even the best judges align with human preference only \textbf{6-18\%} above random chance.
\textbf{TRACE} identifies 8 significantly misaligned rubric items across interaction modalities.
For example, in chat-based coding, judges prefer longer code explanations while humans prefer shorter ones. 
Many of these gaps occur within existing code quality dimensions, showing that current LLM judges remain misaligned with human evaluation in realistic coding workflows.
\end{abstract}

\begin{document}

\maketitle

\section{Introduction}

As LLM-powered tools accelerate software development~\citep{peng2023impactaideveloperproductivity}, there is an increasing need for reliable evaluation methods~\citep{chen2025surveyevaluatinglargelanguage}. 
LLM-as-a-judge has emerged as a widely-used, scalable alternative to human evaluation for assessing model outputs~\citep{li2025generationjudgmentopportunitieschallenges}, including in domains like software engineering~\citep{Wang_2025}. 
Existing work typically considers static settings, using polished code from well-maintained GitHub repositories~\citep{li2024evocodebenchevolvingcodegeneration, 11071936} or competitive programming tasks~\citep{jiang2025codejudgebenchbenchmarkingllmasajudgecoding, qing2025effibenchxmultilanguagebenchmarkmeasuring}. 
While grounded in real software artifacts, these sources fail to capture the messy and underspecified conditions in which developers evaluate and refine code, revealing little about whether LLM judges reflect the implicit criteria developers use in practice.
We ask: \textit{what preferences do LLM judges exhibit when evaluating code, and how do they compare to those of developers?}

\noindent Since real software development rarely occurs in such static settings, evaluation of LLM outputs should capture functional correctness along with developer intent, constraints, and workflow expectations. 
Empirical evidence shows that only 25\% of GitHub autocompletions are accepted by developers and users often report that models fail to meet specific requirements or match their expectations~\citep{ziegler2022productivityassessmentneuralcode,Liang2023ALS}. 
To understand these challenges observed in AI-assisted software development, we examine three representative interaction modalities identified in developer-AI interaction taxonomies~\citep{treude2025developersinteractaitaxonomy}: IDE autocompletion~\citep{chi2025copilotarenaplatformcode}, chat-based programming assistance~\citep{chiang2024chatbotarenaopenplatform}, and instructed code editing~\citep{chi2025editbenchevaluatingllmabilities}. By comparing LLM judgments to developers' preferences across these settings, we analyze how judges weight code quality criteria in practice and where their biases diverge from human preferences.

\begin{figure}[t]
\centering
\includegraphics[width=\textwidth]{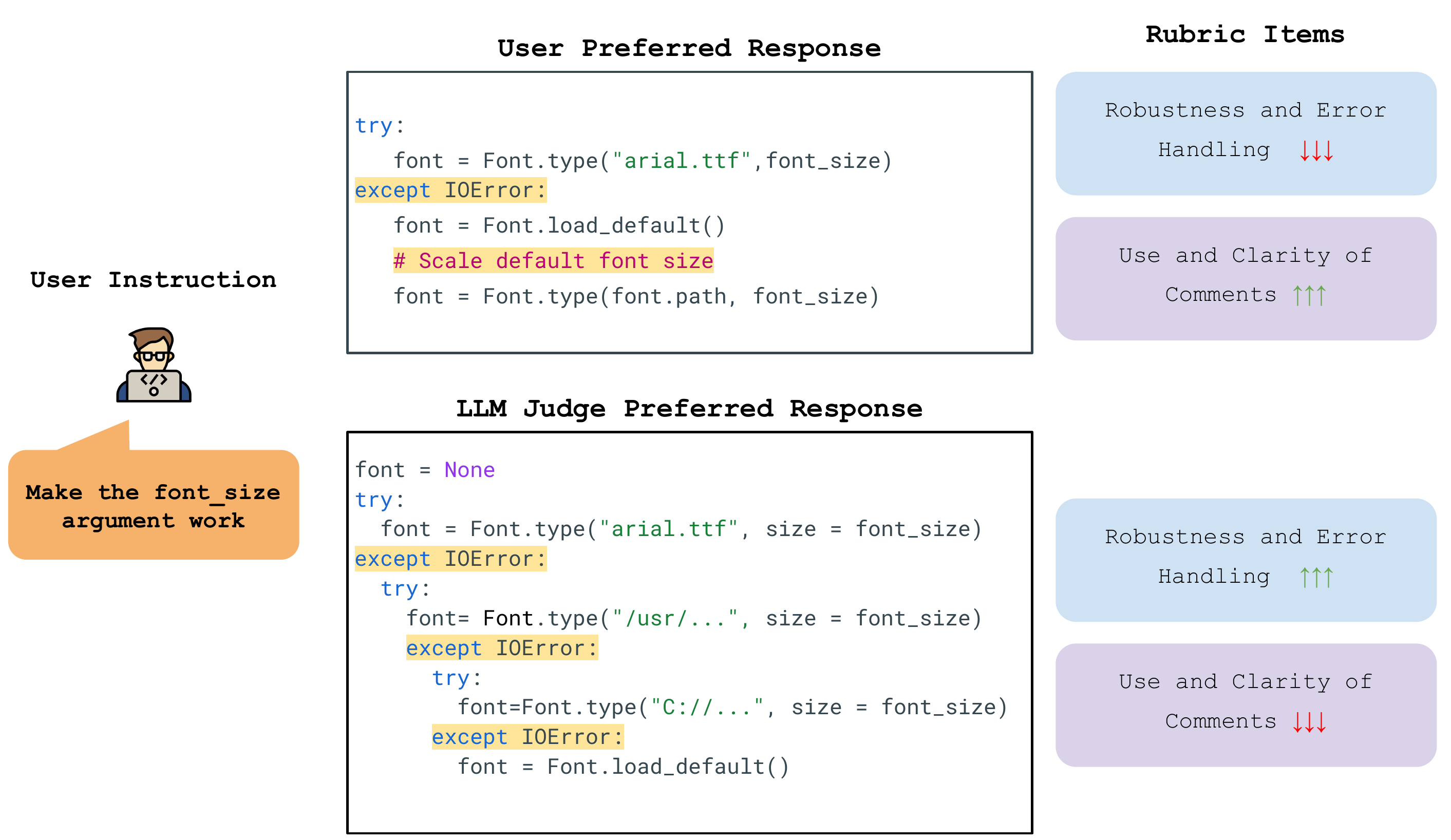}
\caption{\textbf{Example of developer–LLM misalignment on a code editing task.}
Here, the developer provides a prompt and receives two LLM code solutions. In this example, the user prefers the top response while the LLM judge selects the bottom response. We compare these responses with the extracted rubric items to see that the developer
prefers less robustness and more comments, while LLM judges prefer more robustness and fewer comments.
}
\label{fig:motivation}
\vspace{-5pt}
\end{figure}

\noindent We propose \textbf{TRACE} (\textbf{T}ool for \textbf{R}ubric \textbf{A}nalysis in \textbf{C}ode \textbf{E}valuation), the first framework to evaluate and diagnose LLM judge misalignment in realistic coding workflows with interpretable rubric-level analysis.
\textbf{TRACE} measures how closely model judgments align with human preferences in the ambiguity of real-world settings.
Beyond aggregate agreement, we focus on cases of divergence between model and human evaluations. 
To explain these disagreements, \textbf{TRACE} automatically discovers decision criteria that account for judgment differences across samples. Building on prior work in automatic LLM-based criteria discovery~\citep{Dunlap:2025, kim2025evaletevaluatinglargelanguage,findeis2025inverseconstitutionalaicompressing}, we aggregate differences in responses to create a set of qualitative, interpretable ``rubric'' items. 
We then analyze how both human and model judgments correlate with these rubric items, revealing systematic differences in evaluation behavior and preference across judges and modalities (Figure \ref{fig:motivation}). We evaluate across a diverse set of 13 judges, including general-purpose third-party models, specialized judge models, and reward models to show that:

\begin{itemize}[leftmargin=*]
    \item \textbf{LLM judges are consistently misaligned with human code preferences.} Across the three interaction modalities, the strongest LLM judges align with human preference only 6-18\% above random guessing, and no single judge consistently dominates. Notably, fine-tuned judge LLMs do not reliably outperform general-purpose models, suggesting that current shortcomings are not solely due to training data or specialization.

    \item \textbf{In each modality, human and LLM judges are misaligned on different rubric items.} Across the three modalities, 8 rubric items exhibit significant gaps.
    In code completion, LLMs tend to overweight code functionality and underweight readability inside a live file. In edits, judges more often discount clarity, while developers expect changes to be precise. In chat, judges typically reward generic explanations, while humans prefer context-aware solutions. These gaps show there are multiple notions of ``good code'' depending on modality, and judges are unable to reliably capture these nuances.

    \item \textbf{Judges are misaligned on established software engineering criteria.}
    Across three interaction modalities in code, we identify 16 recurring evaluation themes, with 11 aligning closely with canonical software engineering criteria like \textit{syntactic correctness}, \textit{formatting}, and \textit{robustness}. We find that LLM judges remain significantly misaligned with human preference on 6 of these 11 themes across modalities, suggesting that alignment gaps with human preferences on code quality persist during judge training. 
\end{itemize}

\section{Related Work}

\textbf{Interaction Modalities in Software Engineering.} 
LLMs now support a range of interaction modalities
in software development, from real-time, low-overhead code completion~\citep{Svyatkovskiy2020IntelliCodeCC, Pu2025AssistanceOD}, to conversational chat for multi-turn problem solving~\citep{Ross2023ThePA}, to emerging agent systems that autonomously modify codebases~\citep{Chen2025CodeWM, Li2025DeepCodeOA}. Prior work shows these modes induce distinct usage patterns~\citep{Barke2022GroundedCH, Weber2024SignificantPG} and that adoption hinges on reducing effort and accelerating tasks~\citep{Vaithilingam2022ExpectationVE, Liang2023UnderstandingTU, Mozannar2022ReadingBT}, while preserving user control, contextual grounding, and trust~\citep{Chen2025ScreenRP, Liang2023ALS, Brandebusemeyer2025DevelopersExperienceWG, Awad2025PreFilteringCS, Kula2025TheSF, Lyu2025MyPI}. 
Our work studies LLM judges across multiple interaction modalities to evaluate and compare judges with human preferences in software development contexts.

\noindent\textbf{LLM as a Judge.} 
With the growing adoption of LLM judges, researchers have proposed many techniques to align LLMs with human preferences. 
These include fine-tuning approaches~\citep{wang2025djpo}, which produce specialized judge models such as JudgeLM~\citep{zhu2025judgelm}, Prometheus~\citep{kim2024prometheus}, and Atla Selene Mini~\citep{alexandru2025atlaseleneminigeneral}. Additionally, judgment alignment can be improved with inference time methods using multiple LLMs~\citep{verga2024juries} and structured prompting frameworks~\citep{jung2025trust}. 
Specialized benchmarks were developed for systematically comparing LLM judges. 
For example, JudgeBench~\citep{judgebench2024} uses pairwise questions using objective correctness and Arena-Hard~\citep{li2024crowdsourceddatahighqualitybenchmarks} curates high-quality in-the-wild human prompts for evaluation. 
We build on this line of work by evaluating LLM judges in the types of realistic, ambiguous situations developers face in practice.

\noindent\textbf{Explaining LLM Decisions.} 
As LLM judges expand across domains, there is a need to explain LLM judgments~\citep{ryu-etal-2023-retrieval,brake-schaaf-2024-comparing}. 
Existing work like WIMHF~\citep{movva2025whatshumanfeedbacklearning} measures pairwise preferences by training an SAE on embedding differences to encode latent features in responses. Other approaches remove the need for training. VibeCheck~\citep{Dunlap:2025} offers an initial approach to identify evaluation criteria from pairwise differences using LLMs. Evalet~\citep{kim2025evaletevaluatinglargelanguage} provides a method for analyzing LLM judge alignment given a set of evaluation criteria from the user. ICAI~\citep{findeis2025inverseconstitutionalaicompressing} proposes a method to generate explicit instruction criteria for alignment, but these criteria do not generalize across the entire dataset for multiple judges. 
We extend these approaches to automatically discover interpretable rubric items to compare human and LLM judge preference across interaction modalities. 

\section{Methodology}

\begin{figure*}[t]
\centering
\includegraphics[width=\textwidth]{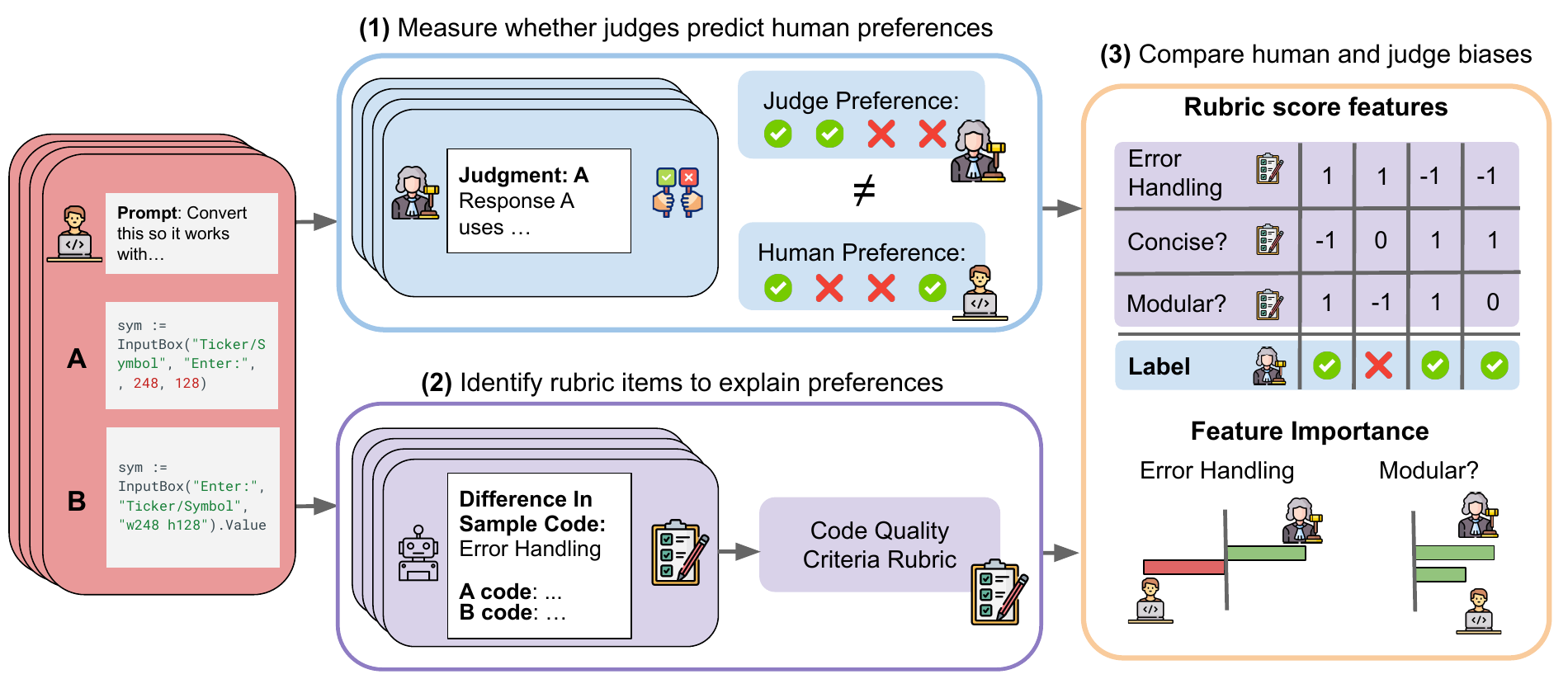}
\caption{\textbf{Overview of TRACE.} Given a set of pairwise options, \textbf{TRACE} follows a three-step workflow: \textbf{(1)} we collect LLM judgments to measure alignment with human preferences; \textbf{(2)} we generate rubric items capturing differences between responses (e.g., \textit{error handling}); \textbf{(3)} we construct feature vectors from rubric criteria and train a model to predict LLM judgments. We use the learned coefficients to identify items that misalign between LLMs and humans.}
\label{fig:overview}
\vspace{-10pt}
\end{figure*}

Given a dataset of human preferences and a set of candidate LLM judges, how do we determine which judge best aligns with human preferences and compare human and judge preferences?
\textbf{TRACE} answers these questions in three stages (Figure \ref{fig:overview}): (1) measure judge agreement with human pairwise preferences, (2) generate interpretable rubric items that capture the evaluative criteria distinguishing response pairs, and (3) compare how humans and judge preferences differ using rubric items.

\noindent\textbf{Step 1: Measure whether judges align with human preference.} Each dataset---see Section~\ref{preference-datasets} for examples---consists of $n$ pairwise preference examples $(x, y_A, y_B, w)$, where $x$ denotes the input context, $y_A$ and $y_B$ are candidate responses, and $w \in \{-1, 1\}$ indicates human preference. For each judge $J$, we provide $(x, y_A, y_B)$ as input and record a binary decision $J(x, y_A, y_B) \in \{-1,1\}$, following prior work~\citep{zheng2023judgingllmasajudgemtbenchchatbot}.
We report both overall accuracy and positionally consistent accuracy. To calculate positionally consistent accuracy, we evaluate each sample twice, once in the original order and once with responses swapped, and discard cases where the judge’s prediction changes. Full prompts are provided in Appendix~\ref{app:predicting-human-preference}.

\noindent\textbf{Step 2: Identify rubric items to explain preferences.}
For each dataset, we construct a rubric $R$ of natural language criteria that characterize differences between pairs of code samples. For instance, a rubric item for \textit{robustness and error handling} may capture whether one response incorporates more exception handling or systematically anticipates edge cases than the other. We create $R$ using two sources: 

\begin{itemize}[leftmargin=*]
    \item \textit{LLM-generated Rubric Items.} 
We generate rubrics using the procedure described in VibeCheck~\citep{Dunlap:2025}.
We sample batches from the dataset and prompt an LLM to describe concrete differences between the two generated responses. 
A subsequent LLM aggregates these differences into a rubric $R_A$ by merging semantically similar items.
\item \textit{Human-Annotated Rubric Items.} To incorporate human judgment into rubric construction, we had annotators review examples from each dataset to create a set of broad evaluative criteria that apply across examples, yielding a rubric $R_H$.

\end{itemize}

\noindent Finally, we aggregate similar items in $R_A$ and $R_H$ using an LLM to produce a final rubric $R$. These rubric items serve as heuristic dimensions for diagnosing differences between judgments rather than definitive criteria for code quality. See Appendix \ref{app:discover-evaluative-criteria} for details.


\noindent\textbf{Step 3: Diagnose human-judge misalignment via rubric-based preference models.}
To characterize disagreements between humans and LLM judges, we treat rubric items in $R$ as a basis for preference decisions and use interpretable models to analyze how humans and LLMs weight each item. 
To do this, we use an LLM to assign a score in $\{-1,0,1\}$ for each sample and rubric item, indicating whether response $y_A$ better satisfies the rubric item ($-1$), $y_B$ does ($1$), or both satisfy it equally ($0$). These scores form a feature matrix $S$, where rows correspond to response pairs and columns correspond to rubric items. 
We then train separate logistic regression models for humans and each LLM judge using $S$ as input features. Human judgments define the labels for the human model, while each judge’s judgments define the labels for its corresponding model. This yields a human preference coefficient vector $\beta_H$ and judge-specific vectors $\beta_J$, where each coefficient captures the influence of rubric item $R_i$ on preferences under the same feature representation.

\noindent We quantify judge-level misalignment using $\beta_{J, i} - \beta_{H, i}$: positive values indicate that a judge places greater emphasis on $R_i$ than humans, while negative values indicate less emphasis. A judge is considered misaligned on rubric $R_i$ when the 95\% confidence interval for $\beta_{J, i}$ excludes $\beta_{H, i}$, with intervals computed via bootstrap resampling. To assess rubric-level misalignment, we pool judge coefficients through a random effects meta-analysis using the Paule-Mandel estimator. This yields a pooled judge coefficient $\hat{\beta}_{J, i}$, and we test whether $\hat{\beta}_{J, i} - \beta_{H, i}$ differs significantly from zero using its standard error; rubric items with significant differences are considered misaligned. See Appendix \ref{app:diagnosing-judge-misalignment} for additional experimental details and robustness checks.


\section{Experimental Set-Up}

Code and implementation details are available in a public repository. \footnote{https://github.com/rShar01/TRACE} The Code Completion and Code Edit data samples were inspected to remove PII, api keys/secrets, and offensive content. Our input set will be released as a part of the public repository.

\subsection{Preference Datasets}
\label{preference-datasets}

We apply \textbf{TRACE} to three representative interaction modalities identified in developer-AI taxonomies~\citep{treude2025developersinteractaitaxonomy}: in-file code completion, instructed code editing, and open-ended chat. Table \ref{tab:dataset-structure} summarizes the differences between these settings.

\begin{table}[t]
\centering
\small
\begin{tabular}{l | r | r | r}
\toprule
 & \textbf{Code Completion} & \textbf{Chat} & \textbf{Code Edit} \\
\midrule
\# of Natural Languages & 23 & 14 & 20 \\
\# of Programming Languages & 39 & 57 & 43 \\
\midrule
Context Length - p50 & 2,233 & 1,013 & 3,490\\
Context Length - p95 & 13,984 & 12,509 & 28,189 \\
\midrule
Output Length - p50 & 108 & 5,491 & 531\\
Output Length - p95 & 613 & 20,156 & 3,053\\
\midrule
Lines of Code - p50 & 78 & 117 & 136\\
Lines of Code - p95 & 383 & 539 & 744\\
\midrule
Edit Distance - p50 & 5 & 123 & 9\\
Edit Distance - p95 & 20 & 579 & 68 \\
\midrule
Natural Language Instruct & & \checkmark & \checkmark \\
Edits Existing Code & \checkmark & & \checkmark \\
In-IDE &\checkmark & & \checkmark  \\
\bottomrule
\end{tabular}
\caption{\textbf{Coding modalities differ sharply in context length and output scale.} 
We summarize three developer interaction settings---code completion, chat, and code edits---each with over 500 samples per dataset.
Natural languages are detected using Lingua \citep{stahl2024lingua}; programming languages in chat are inferred from code block tags. Edit Distance reports the line-level Levenshtein edit distance between the two candidate responses (after newline normalization; for chat, computed on concatenated code blocks when present).}
\label{tab:dataset-structure}
\end{table}

\begin{itemize}[leftmargin=*]
    \item \textbf{Code completions.} We obtain user preferences for code completion from Copilot Arena \citep{chi2025copilotarenaplatformcode}, which collects pairwise judgments through a VSCode extension. The extension presents two fill-in-the-middle completions from different LLMs, and the user selects the one they prefer to insert into their file. We define $x$ as the file context in the workspace, $y_A, y_B$ as the two candidate completions, and $w$ is the user's choice.

    \item \textbf{Chat responses.} Developers commonly interact with LLM using chats. We source pairwise preferences from Chatbot Arena \citep{chiang2024chatbotarenaopenplatform}, where users submit prompts and two LLM assistants generate replies. Since Chatbot Arena is a general-purpose dataset, we filter for 500 code-specific examples (Appendix \ref{app:dataset-filtering}). Here, we define $x$ as the prompt, $y_A, y_B$ as the two chat responses, and $w$ is the user selection.

    \item \textbf{Instructed code edits.} In instructed code edits, users highlight a region of code and provide instructions describing edits. We source these preferences from EDIT-Bench \citep{chi2025editbenchevaluatingllmabilities}, where users select their preferred edit from two LLM solutions. In this setting, $x$ is the user instruction with the highlighted region and file context, $y_A, y_B$ are the two candidate edits, and $w$ is the user selection.
\end{itemize}

\subsection{LLM Judges}
\label{sec:llm-judges}

We consider 13 different candidate LLM judges from the following categories: 

\noindent\textbf{3rd Party Models.}
We include widely deployed general-purpose LLMs, including \texttt{OpenAI GPT-5} \citep{singh2025openaigpt5card}, \texttt{OpenAI GPT-4o} \citep{ahmad2024gpt4o}, \texttt{DeepSeek-R1} \citep{deepseekai2025deepseekr1incentivizingreasoningcapability}, \texttt{Meta Llama-3.1-70B Instruct} \citep{grattafiori2024llama3herdmodels}, and \texttt{Anthropic Claude Sonnet 4} \citep{claude}. We also evaluate smaller variants, such as \texttt{OpenAI GPT-5 mini} \citep{singh2025openaigpt5card} and \texttt{OpenAI o3-mini (high reasoning)} \citep{openai2025o3o4}. These models represent the current frontier of general reasoning systems and serve as baselines for how untuned LLMs perform as judges in code-centric tasks.

\noindent\textbf{Specialized Judge Models.}
We also evaluated models designed specifically for judging. Unlike reward models, these systems produce natural language critiques and decisions, but are trained for evaluation rather than generation. \texttt{Prometheus 2 (7B)} \citep{kim2024prometheus} is a dedicated evaluator that combines scoring and pairwise comparison objectives. \texttt{Atla Selene 1 Mini (8B)} \citep{alexandru2025atlaseleneminigeneral} trains on supervised preferences with a DPO-style ranking loss to sharpen separation between preferred and non-preferred outputs. \texttt{Atla Selene 1 (70B)} scales this design, outperforming frontier models on RewardBench \citep{RewardBench}. \texttt{Skywork Critic (70B)} \citep{skyworkcritic2024} generates synthetic critic data during finetuning and ranks among the leading models on RewardBench. 

\begin{table*}[t]
	\caption{\textbf{Automated judges trail human agreement across coding modalities.} We report accuracy (Acc, \%) and positional accuracy ($\text{Acc}_{\text{PC}}$, \%) across three modalities. $\text{Acc}_{\text{PC}}$ conditions on valid, positionally consistent decisions (the judge flips its choice when the two candidates are swapped). ``Fine-tuned Judge'' models are trained for evaluation and output a discrete winner; ``3rd Party'' models are general-purpose instruction-tuned LLMs used zero-shot as judges; ``Reward Models'' output scalar preference scores and are inherently order-invariant. Human rows report majority--user agreement on a 30-example overlap set and the absolute-point improvement over the best model.}
	\label{tab:all-results}
    \centering
    \small
    \resizebox{\columnwidth}{!}{\begin{tabular}{lcccccc}
    \toprule
    & \multicolumn{2}{c}{Code Completion} & \multicolumn{2}{c}{Instructed Code Edits} & \multicolumn{2}{c}{Chat-based Coding} \\
    & $\text{Acc}_{\text{PC}}$ $\uparrow$ & Acc $\uparrow$ & $\text{Acc}_{\text{PC}}$ $\uparrow$ & Acc $\uparrow$ & $\text{Acc}_{\text{PC}}$ $\uparrow$ & Acc $\uparrow$ \\
    \midrule
    \multicolumn{7}{l}{\textbf{Fine-tuned Judge}} \\
    \midrule
    \texttt{Atla Selene 1 Mini (Llama-3.1-8B)} & 59.01 & 37.67 & 51.8 & 31.6 & 63.76 & 19.00 \\
    \texttt{Atla Selene 1 (Llama-3.3-70B)} & 61.99 & 46.00 & 54.35 & 30.4 & \textbf{67.40} & 24.40 \\
    \texttt{Prometheus 2 (7B)} & 53.62 & 29.60 & 52.21 & 26.0 & 58.80 & 25.40 \\
    \texttt{Skywork Critic (Llama-3.1-70B)} & 63.28 & 48.6 & \textbf{56.14} & 32.00 & 61.87 & 51.6 \\
    \midrule
    \multicolumn{7}{l}{\textbf{3rd Party}} \\
    \midrule
    \texttt{OpenAI GPT-5 mini} & 62.23 & 51.40 & 53.76 & 41.6 & 65.60 & \textbf{57.60} \\
    \texttt{OpenAI GPT-5} & 62.73 & 54.20 & 53.40 & 40.8 & 62.32 & 51.6 \\
    \texttt{OpenAI o3-mini (high reasoning)} & 57.53 & 46.60 & 53.20 & 36.60 & 66.33 & 52.00 \\
    \texttt{OpenAI GPT-4o} & 53.76 & 38.60 & 54.06 & 34.60 & 63.64 & 49.00 \\
    \texttt{Anthropic Claude Sonnet 4} & \textbf{68.14} & 55.60 & 54.04 & 38.8 & 64.53 & 52.40 \\
    \texttt{DeepSeek-R1} & 65.71 & 41.00 & 52.01 & 28.6 & 65.23 & 45.40 \\
    \texttt{Meta Llama-3.1-70B Instruct} & 62.03 & 46.40 & 51.82 & 31.80 & 66.83 & 27.80 \\
    \midrule
    \multicolumn{7}{l}{\textbf{Reward Models}} \\
    \midrule
    \texttt{PairRM} & 50.60 & 50.60 & 47.00 & \textbf{47.00} & 51.80 & 51.80 \\
    \texttt{GRM-Gemma-2B-rewardmodel-ft} & 60.80 & \textbf{60.80} & 45.80 & 45.80 & 53.40 & 53.40 \\
    \midrule
    \multicolumn{7}{l}{\textbf{Human}} \\
    \midrule
    Majority-User Agreement & -- & 83.3 & -- & 66.7 & -- & 70.0 \\
    Annotator Improvement Over Best & -- & 22.5 & -- & 15.9 & -- & 12.4 \\
    \bottomrule
    \end{tabular}}
\end{table*}

\noindent\textbf{Reward Models.} Finally, we evaluate reward models, which output scalar preference scores rather than natural language judgments. \texttt{PairRM} \citep{llm-blender-2023} employs a lightweight pairwise comparison architecture at 0.4B parameters. \texttt{GRM-Gemma-2B-rewardmodel-ft} \citep{yang2024regularizing} derives from \texttt{Gemma-2B} and is fine-tuned on human preference data, reaching state-of-the-art performance for models under 6B parameters on RewardBench.


\section{Results}

\subsection{How well do LLM judges predict human preferences?}

Table \ref{tab:all-results} reports both overall accuracy and positional accuracy for all models across the three modalities. Performance on established judge benchmarks transfers weakly to our modalities, suggesting that benchmarks do not reliably predict real-world preference alignment in interactive coding settings (Appendix \ref{app:benchmark-comparison}).

\noindent\textbf{No model family consistently outperforms the rest.} Third-party frontier models lead in overall accuracy, while fine-tuned judges typically achieve the strongest positional accuracy in the selected modalities. Reward models remain consistently weaker. Overall, differences among fine-tuned, third-party, and reward models remain small, indicating current judge training strategies do not address the sources of misalignment we observe.

\noindent\textbf{Judges exhibit strong position bias.} Across all three datasets, language models exhibit a substantial gap between positional accuracy and overall accuracy. In 3rd-party models, we found that the gap between $\text{Acc}_{\text{PC}}$ and Acc ranges from 8-24\%, and even stronger positional bias exists for fine-tuned judge models with gaps ranging from 10-45\%. Reward models show no such gap because their pairwise scoring is inherently order-invariant. This gap suggests that much of the error arises from sensitivity to input order rather than disagreement alone.

\subsection{How do human and judge preferences compare?}
\label{sec:res-human-judge-bias}

\begin{figure}[t]
\centering
\includegraphics[width=\columnwidth]{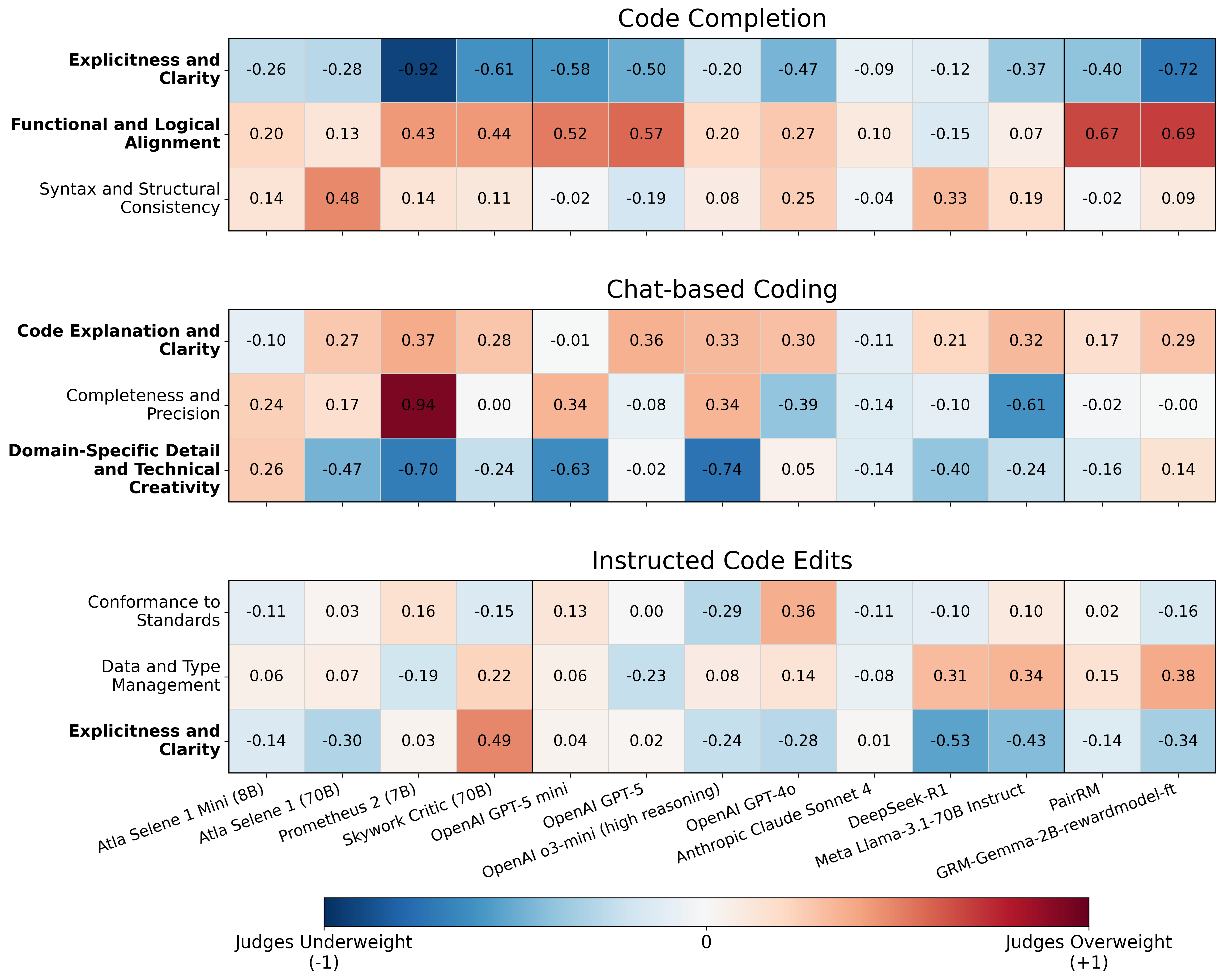}
\caption{\textbf{Judge misalignment highlights distinct rubric-level differences across interaction modalities.} Each cell shows the signed difference between judge and human preference coefficients ($\beta_{J, i} - \beta_{H, i}$) for selected rubric items within each interaction modality. Positive values (red) indicate that judges overweight a rubric item relative to humans, while negative values (blue) indicate underweighting. Rows show the highest-divergence rubric dimensions within each modality. Bolded values indicate rubric items with significant pooled judge misalignment.}

\label{tab:top-misaligned-rubrics}
\centering
\end{figure}

Figure~\ref{tab:top-misaligned-rubrics} highlights rubric-level differences between human and LLM judge preference models across all three modalities. We show the rubric items with the highest divergence for each modality, see Appendix \ref{app:results-judge-misalignment} for all rubric items. 
We discuss insights for each interaction modality:

\noindent\textbf{In code completion settings, judges overvalue the importance of functional code.} Judges systematically underweight \textit{Explicitness and Clarity} and overweight \textit{Functional and Logical Alignment}. This pattern suggests completion judging emphasizes properties that are directly verifiable from the provided code block, such as logical consistency. By contrast, dimensions like clarity may depend on longer-term context, such as team conventions, which users implicitly account for when deciding whether to insert a completion into their codebase. We provide a concrete example of this misalignment in Table \ref{tab:concrete_example}.

\lstset{
  aboveskip=0pt,
  belowskip=0pt,
  language=Java,
  basicstyle=\ttfamily\small,
  breaklines=true,
  columns=fullflexible,
  keepspaces=true,
  showstringspaces=false
}
\begin{table}[t]
\centering
\small
\setlength{\tabcolsep}{6pt}
\begin{tabular}{
    >{\raggedright\arraybackslash}p{3.4cm}
    >{\raggedright\arraybackslash}p{5.5cm}
    >{\raggedright\arraybackslash}p{5.5cm}
}
\toprule
\textbf{Prompt} \newline \textbf{(paraphrased)} &
\multicolumn{2}{p{11cm}}{
Implement the \lstinline{stockSpan} method.
Given an array of daily stock prices, compute, for each day, the number of consecutive previous days whose price is less than or equal to the current price.} \\
\midrule
 & \textbf{Model 1} & \textbf{Model 2} \\
\midrule
\textbf{Completion} &
\begin{minipage}[t]{\linewidth}
\vspace{-6pt}
\begin{lstlisting}
stockSpan(stocks) {
    stack = new Stack<>();
    for (int i = 0; i <
    stocks.length; i++) {
        ...
    }
}
\end{lstlisting}
\end{minipage}
&
\begin{minipage}[t]{\linewidth}
\vspace{-6pt}
\begin{lstlisting}
stockSpan(stocks) {
    stack = new Stack<>({0});
    out[0] = 1;
    for (int i = 1; i < stocks.length; i++) {
        ...
    }
}
\end{lstlisting}
\end{minipage}
\\
\midrule

\textbf{Explicitness} \newline \textbf{and Clarity} &
\begin{minipage}[t]{\linewidth}
{$\boldsymbol{\downarrow}$}\\[5pt]
Omits base case initialization, making initial logic implicit.
\end{minipage}
&
\begin{minipage}[t]{\linewidth}
$\boldsymbol{\uparrow}$ \\[5pt]
States the base case (\lstinline{out[0] = 1}) and starts loop at \lstinline{i = 1}. 
\end{minipage} \\
\midrule

\textbf{Functional and} \newline \textbf{Logical Alignment} &
\begin{minipage}[t]{\linewidth}
$\boldsymbol{\uparrow}$ \\[5pt]
Loop covers all indices without special casing.
\end{minipage}&
\begin{minipage}[t]{\linewidth}
$\boldsymbol{\downarrow}$ \\[5pt]
Adds base case handling and changes loop start, straying from the minimal functional pattern.
\end{minipage}\\
\midrule
\textbf{Judge Preference} & 5 out of 8 & 3 out of 8 \\
\midrule
\textbf{Human Preference} & -- & Selected \\
\bottomrule
\end{tabular}
\caption{\textbf{Example of a rubric item tradeoff in two code completion responses.}
In this example, Model 2 explicitly writes the base case, increasing clarity under the rubric scorer, while Model 1 adopts a more minimal control flow, yielding higher functional alignment. LLM judges more often favor Model 1, while humans select Model 2. Judge counts include only positionally consistent judges.}
\label{tab:concrete_example}
\end{table}

\noindent\textbf{In chat contexts, judges undervalue the importance of domain-aware solutions.} In chat, the dominant gaps shift toward response framing. Judges overweight \textit{Code Explanation and Clarity} and underweight \textit{Domain-Specific Detail and Technical Creativity}. This suggests human users prioritize responses that demonstrate domain awareness and adaptation to their specific problem without excessive explanation. Chat responses provide richer natural language context than code blocks, but judges still appear less sensitive to whether a solution accounts for domain-specific details.

\noindent\textbf{In edits, judges undervalue the importance of clear, unambiguous code.} The same underweighting of \textit{Explicitness and Clarity} appears in edits; this pattern suggests that judges treat edits primarily as constraint satisfaction tasks, focusing on whether the requested change was applied correctly and minimally. Human users, however, appear to treat edits as opportunities to improve code quality, valuing clearer structure and improved readability.

\subsection{How do rubric items map to code quality criteria?}

\begin{table}[t]
\centering
\small
\begin{tabular}{
    >{\centering\arraybackslash}m{2.2cm}
    >{\centering\arraybackslash}m{2.2cm}
    >{\centering\arraybackslash}m{2.2cm}
    >{\centering\arraybackslash}m{2.2cm}
    >{\centering\arraybackslash}m{2.2cm}
    >{\centering\arraybackslash}m{2.2cm}
}
\toprule
\textbf{Shared Across All} & \textbf{Code Edit and Chat} & \textbf{Code Completion and Chat} & \textbf{Code Completion} & \textbf{Code Edit} & \textbf{Chat} \\
\midrule
\cellcolor{blue!10} User-Centeredness & \cellcolor{blue!10} Instruction Following & \cellcolor{blue!10} Creativity / Innovation & \cellcolor{blue!10} Explanatory / Ethical Awareness & \cellcolor{orange!10} Data / Type Management & \cellcolor{blue!10} Domain-Specific Detail \\
\midrule
\cellcolor{orange!10} Conciseness  & \cellcolor{orange!10} Standards / Conventions & \cellcolor{orange!10}Completeness & \cellcolor{orange!10} Syntax / Structural Consistency & & \\
\midrule
\cellcolor{orange!10} Correctness / Precision  & \cellcolor{orange!10} Presentation / Formatting & \cellcolor{orange!10} Efficiency & & \\
\midrule
\cellcolor{orange!10} Modularity / Structure  &  &  & & \\
\midrule
\cellcolor{orange!10} Error Handling / Robustness &  & & & \\
\midrule
\cellcolor{orange!10} Clarity / Explicitness  & & & & \\
\bottomrule
\end{tabular}
\caption{\textbf{A majority of generated rubric items align with existing software engineering criteria.} The orange-highlighted metrics align with established code quality frameworks,
while the blue-highlighted metrics extend beyond traditional software engineering taxonomies,
capturing additional dimensions not typically represented in evaluation criteria. Full rubric groupings appear in Table \ref{table:rubric-generations} and additional comparison with established quality frameworks is present in Appendix \ref{app:results-evaluative-criteria}.} 
\label{table:rubric-overlap}
\end{table}

After generating rubrics independently for each interaction modality (full rubrics in Appendix~\ref{app:results-evaluative-criteria}), we identify semantically similar rubric items across interaction types and cluster them into broader themes (Table \ref{table:rubric-overlap}). We examine how these themes map onto established software engineering criteria and where they extend beyond existing frameworks.

\noindent\textbf{Judge misalignment persists even in established code quality criteria.} 
As shown in Table \ref{table:rubric-overlap}, 11 of 16 themes correspond directly to established software engineering metrics. 
For example, syntax validity is foundational to existing code taxonomies~\citep{Ernst2017WhatTF}, formatting and structural clarity appear in both professional and educational rubrics~\citep{Keuning2023ASM, 10.1145/2999541.2999555}, and conciseness relates to complexity-based measures such as cyclomatic complexity while also capturing notions of minimalism and elegance~\citep{Nilson2019DoIS, AlGhuwairi2023VisualizingSR, Messer2024HowCA}. 
A smaller set of themes has more limited explicit coverage in traditional code quality frameworks, although partial connections exist. Further discussion of these connections is provided in Appendix \ref{app:extends-se-frameworks}.

\noindent Overall, the rubric items generated by our framework reflect the multidimensional view of code quality emphasized in prior literature. Examining model coefficients across these themes shows that 6 of the 11 rubric themes aligned with established software engineering criteria exhibit judge-level misalignment for at least one judge, with 3 remaining significantly misaligned at the rubric level in at least one modality (judge-level results in Appendix \ref{app:stat-sig-judge}). Although many judge models are trained to predict human preferences \citep{ouyang2022traininglanguagemodelsfollow}, they remain misaligned with human judgments along well-established dimensions of software engineering, indicating that substantial alignment gaps persist in these settings.

\section{Conclusions, Limitations, and Future Work}

We presented \textbf{TRACE}, a framework for evaluating LLM judges in realistic code interaction settings and automatically extracting interpretable rubric items that diagnose where model preferences differ from human preference. Across IDE autocompletion, chat-based programming, and instructed code editing, we showed that even the strongest judges remain misaligned with human preferences in realistic developer workflows. Beyond agreement, \textbf{TRACE} identified systematic rubric-level gaps across all modalities, including differences within established code quality dimensions. These results show that improving automated code evaluation will require judge training or calibration methods that explicitly account for rubric-level preference differences.

\noindent\textbf{Limitations and Future Work.} Our framework has several limitations. First, \textbf{TRACE} uses logistic regression to estimate rubric coefficients for interpretability, even though nonlinear models may better capture interactions between rubric dimensions. Future work should test whether more expressive models, paired with explainability methods, improve fidelity. 
Second, while \textbf{TRACE} identifies rubric-level misalignment, it does not provide a mechanism for judge alignment. Future work includes prompt injection with rubric criteria and rubric-conditioned objectives to improve judge models. Third, our analysis only uses 500 preference examples per modality, which results in variability in preference estimates. Despite this, we still observe misalignment across judges and modalities, suggesting these findings reflect systematic trends.

\section*{Acknowledgments}

We thank Bogdan Vasilescu and members of the Sage Lab for their helpful feedback.
This work was supported in part by the National Science Foundation grants IIS1705121, IIS1838017, IIS2046613, IIS2112471, and funding from Datadog. Any opinions, findings and conclusions or recommendations expressed in this material are those of the author(s) and do not necessarily reflect the views of any of these funding agencies.

\bibliography{main}

\appendix
\section{Methodology}

\subsection{Predicting Human Preference}
\label{app:predicting-human-preference}

\subsubsection{Prompt Templates}
These templates define the LLM judge prompts used across our experiments to compare two responses and return a structured preference.

\paragraph{Code Completion Prompt.} 

\begin{quote}
\begin{verbatim}
system: |
  Please act as an impartial judge and evaluate the quality of the 
  responses provided by two AI assistants to the coding context 
  displayed below.

  You should choose the answer that fits the prefix AND suffix 
  contexts within the XML tags. Your evaluation should consider 
  factors such as relevance, accuracy, and style.

  Begin your evaluation by comparing the two responses and provide 
  a short explanation. Avoid any position biases and ensure that 
  the order in which the responses were presented does not 
  influence your decision. Do not allow the length of the 
  responses to influence your evaluation. Do not favor certain 
  names of the assistants. Be as objective as possible.

  After providing your explanation, output your final verdict by 
  strictly following this format within <answer> </answer> XML tags.

  Use the answer "[[A]]" if assistant A is better, "[[B]]" if 
  assistant B is better.

query: |
  <prefix>
  {prefix}
  </prefix>

  <suffix>
  {suffix}
  </suffix>

  <assistant_a_response>
  {answer_a}
  </assistant_a_response>

  <assistant_b_response>
  {answer_b}
  </assistant_b_response>
\end{verbatim}
\end{quote}

\paragraph{Chat-based Interface Prompt.}

\begin{quote}
\begin{verbatim}
system: |
  Please act as an impartial judge and evaluate the quality of 
  the responses provided by two AI assistants to a user prompt.

  The prompt appears below within the XML tag <prompt>, and 
  the two responses appear within tags labelled "Assistant A" 
  and "Assistant B".

  Your evaluation should consider factors such as relevance, 
  accuracy, and style. Begin by comparing the two responses 
  and provide a short explanation.
  Avoid any position biases and ensure the order of 
  presentation does not influence your decision. Do not allow 
  response length to influence your evaluation. Do not favor 
  certain assistant names. Be as objective as possible.

  After providing your explanation, output your final verdict 
  by strictly following this format within <answer> </answer> 
  XML tags.

  Use the answer "[[A]]" if assistant A is better, "[[B]]" if 
  assistant B is better.

query: |
  <prompt>
  {user_instruction}
  </prompt>

  <assistant_a_response>
  {answer_a}
  </assistant_a_response>

  <assistant_b_response>
  {answer_b}
  </assistant_b_response>
\end{verbatim}
\end{quote}

\paragraph{Instructed Code Edit Prompt.}

\begin{quote}
\begin{verbatim}
system: |
  I am a machine learning scientist studying responses given by 
  LLM coding assistants. The models are tasked with editing user 
  code to follow user instructions.

  Please act as an impartial judge and evaluate the quality of 
  the responses provided by the two AI assistants. The responses 
  appear below within XML tags labelled "Assistant A" and 
  "Assistant B".

  Begin your evaluation by comparing the two responses and 
  provide a short explanation. Avoid any position biases and 
  ensure that the order in which the responses were presented 
  does not influence your decision. Do not allow the length of 
  the responses to influence your evaluation. Do not favor
  certain names of the assistants. Be as objective as possible.

  After providing your explanation, output your final verdict 
  by strictly following this format within <answer> </answer> 
  XML tags.

  Use the answer "[[A]]" if assistant A is better, "[[B]]" if 
  assistant B is better.

query: |
  This is the prefix of the coding file:
  {prefix}

  This is the suffix of the file:
  {suffix}

  This is the code selected by the user to rewrite:
  {code_to_edit}

  The user has given the instructions:
  {user_input}

  Below are the assistant-generated edits to the code:

  <assistant_a_response>
  {answer_a}
  </assistant_a_response>

  <assistant_b_response>
  {answer_b}
  </assistant_b_response>
\end{verbatim}
\end{quote}

\subsubsection{Reward Models}
\label{app:reward-models}

Reward models assign scalar scores to LLM outputs to indicate their alignment with human preferences. To run inference for a reward model $J$, we evaluate the input $x$ paired with each candidate response, $(x, y_A)$ and $(x, y_B)$, independently, yielding scores $s_A$ and $s_B$. The model preference is defined as $J(x, y_A, y_B) = -1$ when $s_A < s_B$, and $1$ otherwise. Many modern reward models adapt this framework (see Section \ref{sec:llm-judges} for examples).

\subsection{Discovering Evaluative Criteria}
\label{app:discover-evaluative-criteria}

\paragraph{Configuration.} In our pipeline, \texttt{GPT-4o} is the rubric proposer. For efficiency, we process thirty samples in batches of five during each generation pass. We repeat this step three times, for a total of ninety samples, to produce a more complete rubric set. For rubric proposal and aggregation, we follow the same prompt structure and parameters used in VibeCheck. 

\noindent To derive human-annotated rubric items, we had three engineers review 30 samples from each dataset and write detailed explanations for pairwise preference selections. We aggregated the engineers' rationales for each dataset using an LLM aggregation prompt to create general criteria that are applicable for all samples in the dataset. 

\subsubsection{Prompt Templates}

\paragraph{Proposer Prompt.} This prompt adapts the original VibeCheck proposer template to generate evaluative criteria, with modifications to produce more specific and unique rubric items.

\begin{quote}
\begin{verbatim}
You are a machine learning researcher analyzing two large language 
models (LLMs) by comparing how their responses differ to the same 
set of questions. Your goal is to identify unique, interpretable 
behavioral dimensions ("axes of variation") that capture subtle or 
surprising differences between the models.

Here are the questions and responses:
{combined_responses}

For each axis, describe what makes one model's responses higher 
and the other's lower on that dimension. Focus on differences 
that reveal deeper behavioral tendencies rather than surface 
traits.

Format your output as a bulleted list, with each axis on a new 
line starting with a dash (-) or asterisk (*). Each axis should 
follow this format:

- {axis}: High → {description of high end} | Low → {description 
    of low end}

Example:
- Self-consistency: High → Responses maintain consistent 
reasoning throughout | Low → Reasoning may shift or contradict 
earlier statements

Guidelines:
- Avoid obvious or generic dimensions such as "clarity," 
    "conciseness," or "formality."
- Look for behavioral nuances from reasoning patterns, goal 
    orientation, implicit assumptions, moral framing, creativity 
    style, uncertainty handling, or tone of confidence.
- Axes may mix abstract and domain-specific aspects.
- Each axis must be something a human could use to categorize 
    which model response is higher or lower.
- Do not add explanations, prefaces, or summaries.
- If no substantive differences exist, output only "No 
    differences found."
\end{verbatim}
\end{quote}

\paragraph{Aggregation Prompt.} This prompt adapts the VibeCheck reduction template to aggregate rubric items, prioritizing unique, task-specific dimensions over very general criteria.

\begin{quote}
\begin{verbatim}
The following are axes of variation for comparing two model outputs. 
Each axis includes a name and a description of what makes an output 
high or low on that dimension. Some axes may be redundant, misnamed, 
or overlap with others. Your task is to cluster and reduce these 
axes into a minimal set of parent axes that are as distinct and 
non-overlapping as possible, while preserving the specificity
and uniqueness of the original axes. Do not over-merge genuinely 
distinct properties.

For each parent axis you create:
- Ensure the high and low descriptions faithfully subsume the axes 
    they replace, while retaining distinctive properties rather 
    than over-generalizing.
- If an axis is truly unique or nuanced, keep it as its own parent 
    axis rather than forcing a merge.
- Parent axes must be mutually exclusive and enable a human to 
    reliably and uniquely categorize model outputs along each 
    dimension.
- If an axis is domain- or task-specific (e.g., coding), reflect 
    this specificity in the axis name.

Here are the axes of variation (each formatted as
{axis name}: High: {high description} Low: {low description}):

{differences}

Cluster and reduce these axes into a minimal, clear set of parent 
axes, retaining uniqueness where present. Each parent axis should 
include a name and a concise (<20 words) description that preserves 
any domain-specific or distinctive properties in the original.

Format your output as a bulleted list, one axis per line, using:

- {axis}: High → {description of high end} | Low → {description 
of low end}
\end{verbatim}
\end{quote}

\paragraph{Annotator-Comment Proposer Prompt.}
\label{app:annotator-comment-proposer-prompt}

This prompt identifies evaluative criteria from annotator comments, discovering criteria that reflect how humans differentiate between two candidate responses.

\begin{quote}
\begin{verbatim}
You are a machine learning researcher analyzing annotator comments 
to surface unique, interpretable behavioral dimensions ("axes of 
variation") that capture what annotators notice when preferring one 
answer over another. Work only from the comments -- do not assume 
anything about the original questions or answers.

Here are the comments to analyze:
{comments}

For each axis, describe what makes a response higher versus lower 
on that dimension. Focus on differences that reveal deeper 
behavioral tendencies rather than surface traits.

Format your output as a bulleted list, with each axis on a new line 
starting with a dash (-) or asterisk (*). Each axis should follow 
this format:

- {axis}: High → {description of high end} | Low → {description 
of low end}

Guidelines:
- Derive axes only from the themes present in the comments (e.g., 
    syntax validity, conciseness, unnecessary extras, instruction 
    alignment).
- Look for interpretable, discriminative properties (reasoning 
    patterns, goal orientation, adherence to constraints) rather 
    than generic "good/bad."
- Keep axes human-usable; a reviewer should be able to place an 
    answer as higher or lower on the axis from the comment.
- Do not mention specific questions, models, or options -- focus 
    on underlying properties.
- If no substantive differences are present, output only "No 
    differences found."
\end{verbatim}
\end{quote}

\subsection{Diagnosing Judge Misalignment} 
\label{app:diagnosing-judge-misalignment}
\subsubsection{Rubric Scoring}

\paragraph{Configuration.} In our pipeline, \texttt{GPT-5.1} serves as the rubric scorer. For efficiency, we evaluate five rubric axes per sample in a single scoring pass. We retry malformed outputs up to three times and assign a neutral score when a value remains missing. We follow the VibeCheck scoring setup, including its prompt template and parameters, for this stage of the pipeline.

\paragraph{Positional Bias.} To control for positional bias, we evaluate every pair twice, once in the original order and once with the responses swapped. We retain rubric scores only when the scorer is positionally consistent, meaning the preference flips under swapping. For inconsistent cases, we set the corresponding rubric score to neutral.

\paragraph{Testing Consistency.} Since our pipeline relies on stable LLM based rubric scoring, we repeated the scoring process twice and evaluated agreement between runs. The scorer remained highly consistent across repeated runs (correlation 0.9052), suggesting downstream preference coefficients are not affected by scorer variability.

\section{Experimental Setup}

\subsection{Dataset Filtering and Normalization}
\label{app:dataset-filtering}

We apply dataset-specific preprocessing to ensure that each instance corresponds to a well-formed pairwise preference example of the form $(x, y_A, y_B, w)$ with sufficient context to evaluate the responses. Across all datasets, we (i) require an explicit human preference between two candidates (no ties), (ii) drop rows with malformed serialization or missing required fields, and (iii) remove degenerate pairs where the two candidates are identical when applicable.
We encode preferences using $w \in \{-1, 1\}$, where $w=1$ indicates that the user preferred $y_A$ and $w=-1$ indicates that the user preferred $y_B$.

\subsubsection{Copilot Arena (Code Completion)}
Copilot Arena logs come from a VSCode extension that presents two fill-in-the-middle code completions for the same cursor context. Each record includes the surrounding code context (preceding and following text) and two candidate completions. We parse the serialized completion metadata and retain only examples with a valid user preference between the two candidates. In our implementation, we keep only records where the user accepted the \textit{second} of the two presented completions, which mitigates the possibility that users accept the first completion with minimal comparison and never meaningfully inspect the alternative. Under this filtering, the preferred completion always corresponds to the second candidate, so we set $w=-1$. We additionally require that the completion metadata contains the preceding code context.

\subsubsection{LMArena (Chat)}
For chat-based assistance, we use the \texttt{lmarena-ai/arena-human-preference-140k} dataset. Because LMArena covers a wide range of domains, we apply additional constraints to isolate code-centric, comparable examples. We retain only instances annotated as code-related with a decisive preference for one of the two candidates, and we restrict to the canonical presentation order used for evaluation. To reduce variation due to conversational history, we further require single-turn conversations for both candidates (exactly one user message and one assistant response).

\noindent To focus on developer-like completion/edit interactions, we additionally filter to prompts that match an \textit{edit-like} heuristic (e.g., containing common edit/repair verbs or placeholder markers) while excluding prompts that are primarily explanatory (e.g., definition or “explain” requests) or unrelated to code editing (e.g., image-generation requests).

\noindent Finally, we require that \textit{both} assistant responses contain fenced code blocks (triple backticks) with language tags drawn from a curated set of programming-language identifiers, and that the two responses share at least one such language tag. We also require code-like tokens within the fenced regions to remove prose-only fences. Together, these constraints remove comparisons where candidates respond in different programming languages or where one response is not substantively code.

\noindent To construct an LMArena subset that is comparable to the completion and edit datasets, we further restrict the filtered set to a list of manually retained question identifiers. This list is created by three annotators via inspection of the prompt--response traces, retaining only examples that match our intended coding interaction and contain sufficient context for a meaningful preference judgment.

\subsubsection{EDIT-Bench (Code Edits)}
EDIT-Bench examples consist of a natural-language edit instruction, a code span to edit, and file context (preceding and following text), paired with two candidate edits. The raw CSV stores the candidate data in a string-serialized field; we safely parse this field and drop rows that fail to deserialize. We then restrict to research-consented examples with a binary preference label over the two candidates. We keep only rows that contain the required context (instruction, code span, and file context) and extract both candidate edits to form $(x, y_A, y_B, w)$. To avoid trivial comparisons, we remove pairs where the two candidate edits are identical after trimming leading/trailing whitespace. The winner label $w$ is taken from the recorded preference (first candidate preferred $\rightarrow w=1$; second candidate preferred $\rightarrow w=-1$).

\section{Results}

\subsection{Comparison to other benchmarks}
\label{app:benchmark-comparison}
Across models, performance on our modalities aligns only weakly with established judge benchmarks (Table \ref{tab:cross-benchmark}). Strong results on external judge tasks do not necessarily carry over. \texttt{Skywork Critic (70B)} leads on RewardBench yet ranks only mid-range on code completion and code edits. \texttt{OpenAI GPT-4o} tops JudgeBench (Coding) but shows uneven performance across interaction modalities.

\begin{table*}[t]
\centering
\small
\setlength{\tabcolsep}{3pt}
\resizebox{\columnwidth}{!}{%
\begin{tabular}{
    >{\centering\arraybackslash}m{3.2cm}
    >{\centering\arraybackslash}m{2.5cm}
    >{\centering\arraybackslash}m{1.3cm}
    >{\centering\arraybackslash}m{1.3cm}
    >{\centering\arraybackslash}m{1.8cm}
    >{\centering\arraybackslash}m{1.8cm}
    >{\centering\arraybackslash}m{2.5cm}
}
\toprule
& \multicolumn{3}{c}{Interaction Modalities} & \multicolumn{3}{c}{External  Benchmarks} \\
\cmidrule(lr){2-4} \cmidrule(lr){5-7}
\textbf{Model} & \textbf{Code Completion} & \textbf{Chat} & \textbf{Code\allowbreak Edit} & \textbf{Reward\allowbreak Bench} & \textbf{Reward\allowbreak Bench 2} & \textbf{JudgeBench (Coding)} \\
\midrule
\texttt{Skywork Critic} & 48.6 & 51.6 & 32.0 & 93.3 & -- & 47.6 \\
\texttt{GPT-5 mini} & 45.9 & 32.4 & 32.0 & 80.1 & 58.0 & 45.2 \\
\texttt{GPT-4o} & 38.6 & 49.0 & 34.6 & 86.7 & 64.9 & 59.5  \\
\texttt{Claude Sonnet 4} & 55.6 & 52.4 & 38.8 & -- & 71.2 & -- \\
\texttt{Llama-3.1-70B} & 46.4 & 27.8 & 31.8 & 84.0 & -- & -- \\
\texttt{GRM-Gemma-2B} & 60.8 & 53.4 & 45.8 & -- & 59.7 & 54.8\\
\bottomrule
\end{tabular}
}
\caption{\textbf{External judge benchmarks correlate weakly with completion/edit accuracy.} We report model accuracy (Acc, \%) on our three interaction modalities—IDE code completion, chat-based coding, and instructed code edits—alongside each model’s published score (\%, higher is better) on established judge benchmarks (RewardBench, RewardBench 2, and JudgeBench-Coding). Dashes indicate unavailable results. Models are abbreviated for space.}
\label{tab:cross-benchmark}
\end{table*}

\subsection{Evaluating LLM Judges}
\subsubsection{Controlling for Context Length}
\label{app:context-length}
Table \ref{tab:context-length-results} shows a consistent gap between full-prompt and truncated-prompt performance across all three benchmarks. Models achieve higher accuracy when the entire prompt fits within the model’s context window. Performance degrades when examples exceed that window and the prompt must be truncated. The decline is often substantial for models with shorter context windows, while models with larger context windows exhibit smaller drops. Regardless, even when we restrict evaluation to examples whose full prompt fits within the model’s context window, these models largely do not achieve accuracy comparable to the top model previously reported in each benchmark.

\begin{table*}[t]
	    \centering
	    \small
        \resizebox{\columnwidth}{!}{%
	    \begin{tabular}{lccccccc}
    \toprule
    & \multicolumn{1}{c}{} & \multicolumn{2}{c}{Code Completion} & \multicolumn{2}{c}{Chat} & \multicolumn{2}{c}{Code Edit} \\
    & Context Length & $\text{Acc}_{\text{F}}$ $\uparrow$ & $\text{Acc}_{\text{T}}$ $\uparrow$ & $\text{Acc}_{\text{F}}$ $\uparrow$ & $\text{Acc}_{\text{T}}$ $\uparrow$ & $\text{Acc}_{\text{F}}$ $\uparrow$ & $\text{Acc}_{\text{T}}$ $\uparrow$ \\
    \midrule
    \multicolumn{7}{l}{\textbf{Fine-tuned Judge}} \\
    \midrule
    \texttt{Atla Selene 1 Mini} & 2048 & 42.22 & 19.01 & 29.70 & 16.29 & 36.17 & 25.69 \\
    \texttt{Atla Selene 1 } & 4096 & 47.28 & 18.18 & 33.33 & 22.11 & 36.64 & 34.58 \\
    \texttt{Prometheus 2 (7B)} & 2048 & 33.53 & 21.25 & 37.04 & 23.99 & 28.28 & 23.83 \\
    \texttt{Skywork Critic } & 4096 & 49.37 & 31.82 & 52.09 & 51.05 & 33.84 & 25.23  \\
    \midrule
    \multicolumn{7}{l}{\textbf{Reward Models}} \\
    \midrule
    \texttt{PairRM} & 2048 & 53.77 & 48.5 & 52.63 & 50.7 & 50.00 & 45.76  \\
    \texttt{GRM-Gemma-2B} & 3000 & 54.63 & 100.0 & 46.15 & 53.80 & 47.2 & 42.86 \\
    \midrule 
	    Top Model & \textminus & 60.80 & 60.80 & 57.60 & 57.60 & 47.00 & 47.00 \\
	    \bottomrule
	    \end{tabular}%
        }
    \caption{\textbf{Judging accuracy drops when prompts exceed a model’s context window.} The full-prompt accuracy ($\text{Acc}_{\text{F}}$) is computed on examples whose prompt fits within the model’s context length. The truncated-prompt accuracy ($\text{Acc}_{\text{T}}$) is computed on examples that require truncation. Models are abbreviated for space.
    }
    \label{tab:context-length-results}
\end{table*}

\subsection{Identifying Judge Misalignment}
\label{app:results-judge-misalignment}

\begin{figure*}[t]
\centering
\includegraphics[width=\linewidth]{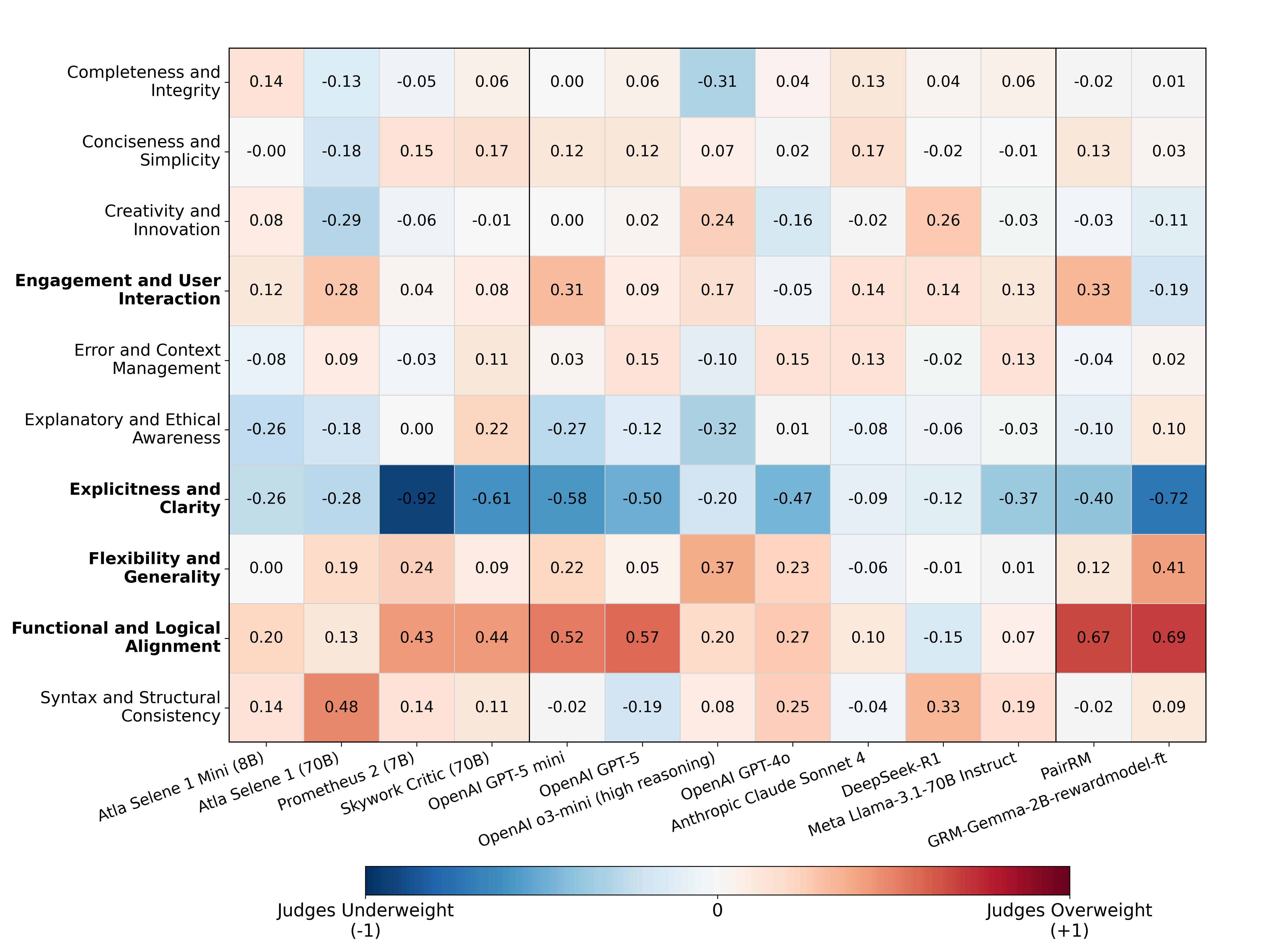}
\caption{
Code completion alignment across all judges and rubrics.}
\label{fig:copilot-heatmap}
\end{figure*}

\begin{figure}[t]
\centering
\includegraphics[width=\linewidth]{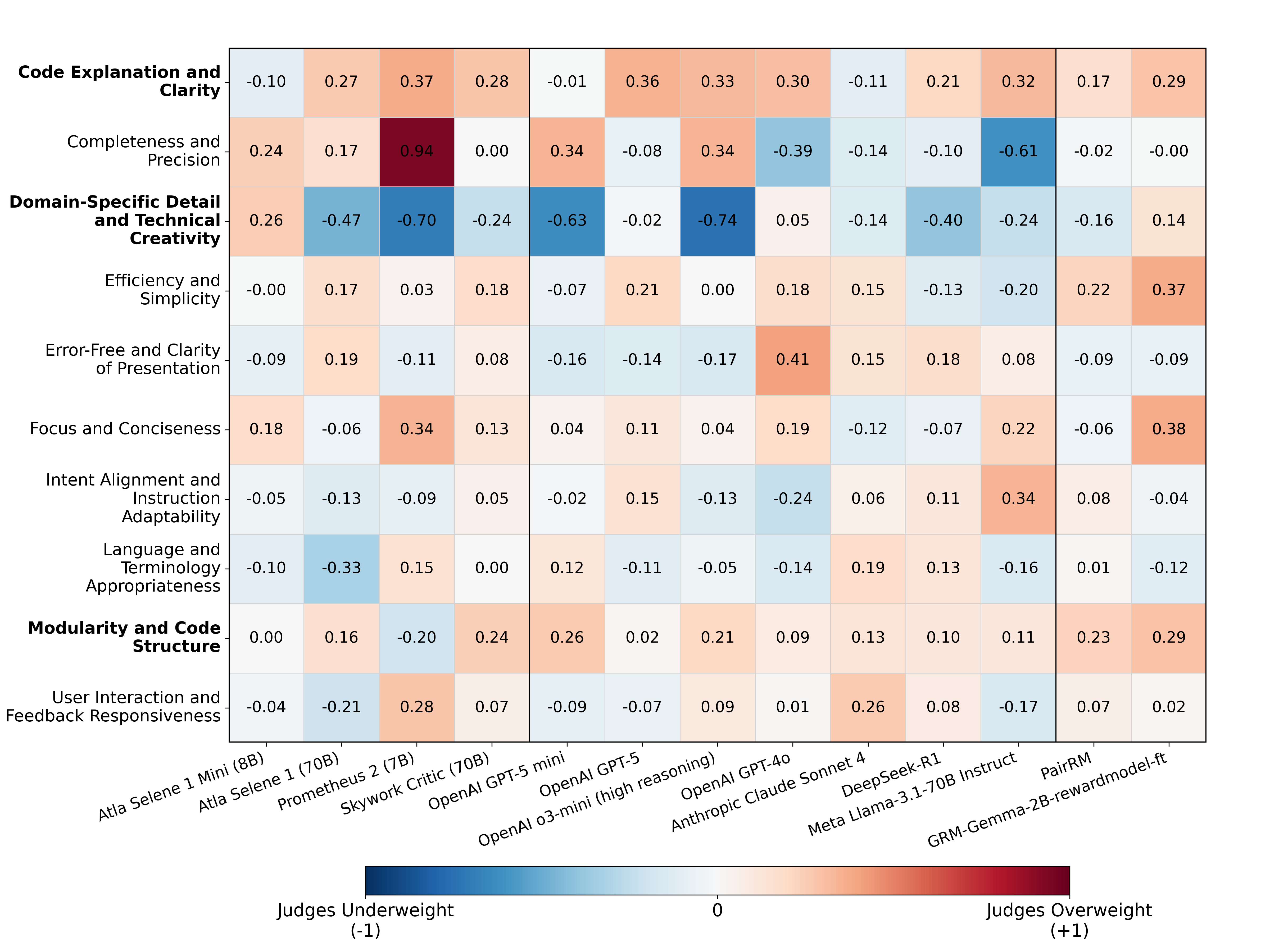}
\caption{
Chat completion alignment across all judges and rubrics.
}
\label{fig:lmarena-heatmap}
\end{figure}

\begin{figure}[t]
\centering
\includegraphics[width=\linewidth]{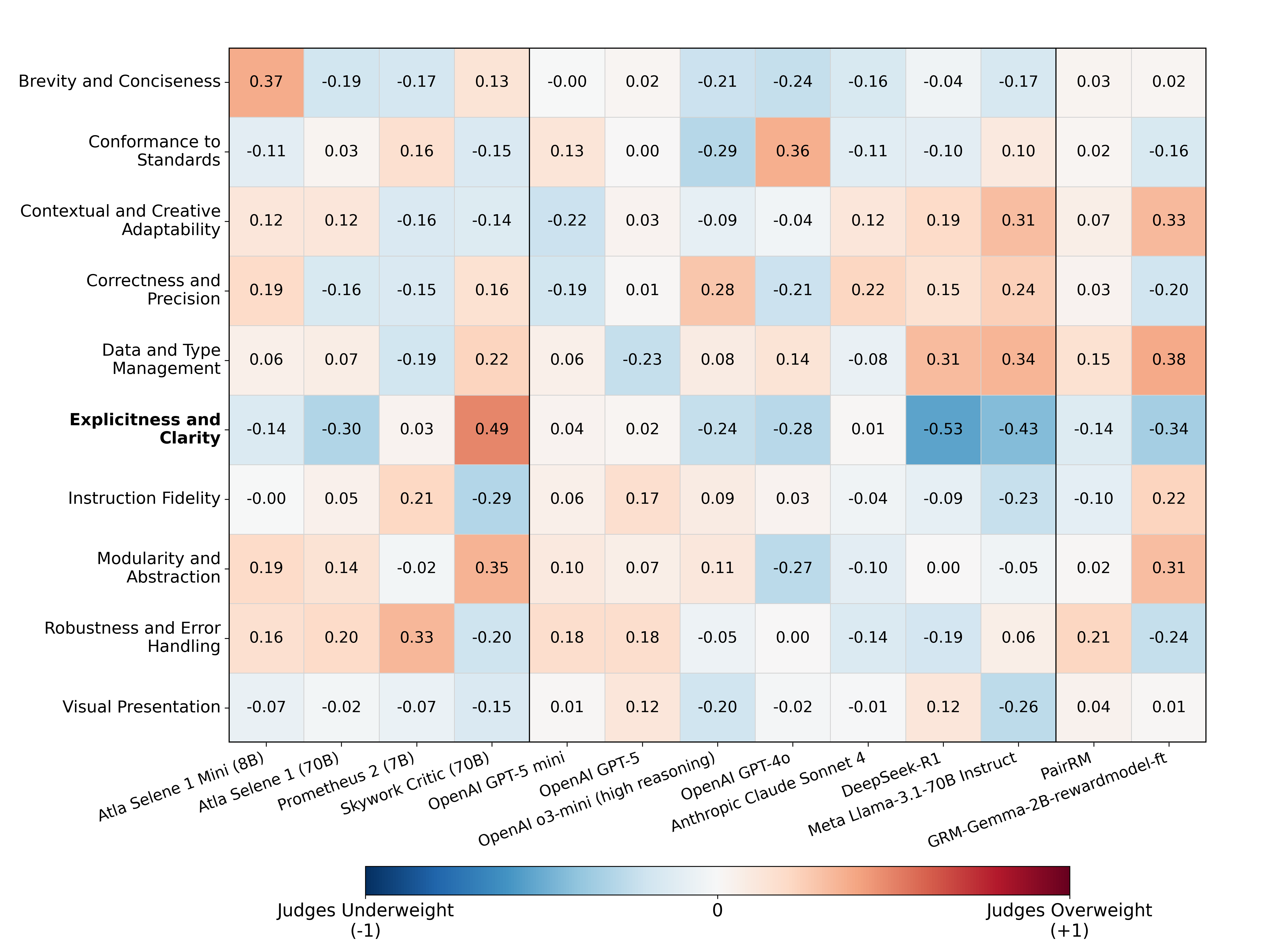}
\caption{
Code edit alignment across all judges and rubrics.}
\label{fig:editbench-heatmap}
\end{figure}

Figures \ref{fig:copilot-heatmap}, \ref{fig:lmarena-heatmap}, and \ref{fig:editbench-heatmap} analyze the signed coefficient difference $\beta_{J, i} - \beta_{H, i}$ for each judge and rubric item $R^{(i)}$ across the code completion, code edit, and chat modalities respectively. We bold significant judge-human gaps, defined when the 95\% confidence interval for $\beta_{J, i}$ excludes $\beta_{H, i}$. 

\subsubsection{Statistically Significant Judge-Rubric Gaps}
\label{app:stat-sig-judge}
Table~\ref{tab:significant-pairs} lists all judge--rubric pairs where the judge’s 95\% CI for $\beta_{J, i}$ excludes $\beta_{H, i}$, indicating a statistically detectable difference in how that judge weights the rubric item relative to humans. We identify 35 significant sources of misalignment between humans and judges across interaction modalities. 

\begin{table}[t]
\centering
\scriptsize
\setlength{\tabcolsep}{4pt}
\begin{tabular}{p{5.4cm} p{6cm} c r}
\toprule
\textbf{Judge} & \textbf{Rubric} & \textbf{Dir} & \textbf{$\Delta = \beta_J - \beta_H$} \\
\midrule
\multicolumn{4}{l}{\textbf{Code Completion}} \\
\midrule
\texttt{Prometheus 2 (7B)} & \textit{Explicitness and Clarity} & $\downarrow$ & -0.924 \\
\texttt{GRM-Gemma-2B-rewardmodel-ft} & \textit{Explicitness and Clarity} & $\downarrow$ & -0.721 \\
\texttt{GRM-Gemma-2B-rewardmodel-ft} & \textit{Functional and Logical Alignment} & $\uparrow$ & +0.691 \\
\texttt{PairRM} & \textit{Functional and Logical Alignment} & $\uparrow$ & +0.669 \\
\texttt{Skywork Critic (70B)} & \textit{Explicitness and Clarity} & $\downarrow$ & -0.605 \\
\texttt{OpenAI GPT-5 mini} & \textit{Explicitness and Clarity} & $\downarrow$ & -0.578 \\
\texttt{OpenAI GPT-5} & \textit{Functional and Logical Alignment} & $\uparrow$ & +0.574 \\
\texttt{OpenAI GPT-5 mini} & \textit{Functional and Logical Alignment} & $\uparrow$ & +0.520 \\
\texttt{OpenAI GPT-5} & \textit{Explicitness and Clarity} & $\downarrow$ & -0.500 \\
\texttt{Atla Selene 1 (70B)} & \textit{Syntax and Structural Consistency} & $\uparrow$ & +0.482 \\
\texttt{GRM-Gemma-2B-rewardmodel-ft} & \textit{Flexibility and Generality} & $\uparrow$ & +0.414 \\
\texttt{OpenAI o3-mini (high reasoning)} & \textit{Flexibility and Generality} & $\uparrow$ & +0.374 \\
\texttt{PairRM} & \textit{Engagement and User Interaction} & $\uparrow$ & +0.331 \\
\texttt{OpenAI o3-mini (high reasoning)} & \textit{Explanatory and Ethical Awareness} & $\downarrow$ & -0.317 \\
\texttt{OpenAI GPT-5 mini} & \textit{Engagement and User Interaction} & $\uparrow$ & +0.312 \\
\texttt{Atla Selene 1 (70B)} & \textit{Creativity and Innovation} & $\downarrow$ & -0.291 \\
\midrule
\multicolumn{4}{l}{\textbf{Chat}} \\
\midrule
\texttt{Prometheus 2 (7B)} & \textit{Completeness and Precision} & $\uparrow$ & +0.939 \\
\texttt{OpenAI o3-mini (high reasoning)} & \textit{Domain-Specific Detail and Technical Creativity} & $\downarrow$ & -0.740 \\
\texttt{Prometheus 2 (7B)} & \textit{Domain-Specific Detail and Technical Creativity} & $\downarrow$ & -0.696 \\
\texttt{OpenAI GPT-5 mini} & \textit{Domain-Specific Detail and Technical Creativity} & $\downarrow$ & -0.625 \\
\texttt{Meta Llama-3.1-70B Instruct} & \textit{Completeness and Precision} & $\downarrow$ & -0.606 \\
\texttt{OpenAI GPT-4o} & \textit{Error-Free and Clarity of Presentation} & $\uparrow$ & +0.410 \\
\texttt{OpenAI GPT-4o} & \textit{Completeness and Precision} & $\downarrow$ & -0.394 \\
\texttt{OpenAI GPT-5} & \textit{Code Explanation and Clarity} & $\uparrow$ & +0.355 \\
\texttt{OpenAI o3-mini (high reasoning)} & \textit{Code Explanation and Clarity} & $\uparrow$ & +0.328 \\
\texttt{OpenAI GPT-4o} & \textit{Code Explanation and Clarity} & $\uparrow$ & +0.303 \\
\texttt{Skywork Critic (70B)} & \textit{Code Explanation and Clarity} & $\uparrow$ & +0.280 \\
\texttt{Anthropic Claude Sonnet 4} & \textit{User Interaction and Feedback Responsiveness} & $\uparrow$ & +0.259 \\
\texttt{Skywork Critic (70B)} & \textit{Modularity and Code Structure} & $\uparrow$ & +0.237 \\
\midrule
\multicolumn{4}{l}{\textbf{Code Edit}} \\
\midrule
\texttt{DeepSeek-R1} & \textit{Explicitness and Clarity} & $\downarrow$ & -0.534 \\
\texttt{Skywork Critic (70B)} & \textit{Explicitness and Clarity} & $\uparrow$ & +0.489 \\
\texttt{Meta Llama-3.1-70B Instruct} & \textit{Explicitness and Clarity} & $\downarrow$ & -0.431 \\
\texttt{GRM-Gemma-2B-rewardmodel-ft} & \textit{Data and Type Management} & $\uparrow$ & +0.379 \\
\texttt{OpenAI GPT-4o} & \textit{Conformance to Standards} & $\uparrow$ & +0.364 \\
\texttt{Skywork Critic (70B)} & \textit{Modularity and Abstraction} & $\uparrow$ & +0.347 \\
\bottomrule
\end{tabular}
\caption{\textbf{Statistically detectable gaps cluster in a small number of rubric dimensions.} We consider a judge--rubric pair to be significant if the judge’s 95\% CI for $\beta_{J, i}$ excludes $\beta_{H, i}$. The direction column indicates whether the judge overweights ($\uparrow$) or underweights ($\downarrow$) the rubric item relative to humans.}
\label{tab:significant-pairs}
\end{table}

\subsection{Discovering Evaluative Criteria}
\label{app:results-evaluative-criteria}

Tables \ref{tab:copilot-rubric-items}, \ref{tab:lmarena-rubric-items}, and \ref{tab:editbench-rubric-items} list the evaluative criteria produced by our rubric construction pipeline for the code completion, code edit, and chat modalities respectively. For each dataset, we report the final set of rubric axes, along with short descriptions of the upper and lower ends of each axis. We also include minimal code examples for each axis to make these rubrics more concrete. Table~\ref{table:rubric-generations} summarizes how these rubric axes align across datasets.

\paragraph{Code quality criteria shift across interaction modalities.} 
The modality-specific columns in Table \ref{table:rubric-overlap} show how different interaction settings give rise to distinct evaluation criteria. 
Code completion rubric items emphasize low-level program concerns like \textit{Syntax/Structural Consistency}, reflecting the need for a completion to fit seamlessly into an existing file.
Rubric items for code edits mainly capture compliance criteria such as \textit{Data/Type Management}, where judges assess whether a model obeys explicit instructions and code invariants. 
Chat-based criteria emphasize communication behavior through \textit{Domain-Specific Detail}, where responses provide reasoning and explanations beyond code edits.

\paragraph{Rubric items connect to traditional code quality methods.}
\label{app:traditional-swe-metrics}
Several elements are closely related to well-studied quality dimensions. \textit{Syntax/Structural Consistency}, \textit{Presentation/Formatting}, \textit{Conciseness}, and \textit{Correctness/Precision} align with core software quality principles. Syntax validity is foundational to existing code taxonomies~\citep{Ernst2017WhatTF}, formatting and structural clarity appear in both professional and educational rubrics~\citep{Keuning2023ASM, 10.1145/2999541.2999555}, and conciseness relates to complexity-based measures such as cyclomatic complexity while also capturing notions of minimalism and elegance~\citep{Nilson2019DoIS, AlGhuwairi2023VisualizingSR, Messer2024HowCA}. Functional and logical alignment reflects functional and methodological correctness \citep{Messer2023AutomatedGA}. \textit{Error Handling/Robustness} corresponds to reliability-focused evaluation criteria~\citep{Bishop2024EvaluatingSC} and engineering-oriented performance metrics~\citep{Hariharan2025SemanticME}. More granular criteria, such as \textit{Data/Type Management}, emphasize type safety and error prevention \citep{hanenberg2013}. Additional themes—\textit{Clarity/Explicitness}, \textit{Modularity/Structure}, and \textit{Completeness}—map to established notions of readability, modularization, design quality, and problem coverage \citep{Keuning2023ASM, Ernst2017WhatTF, Tablan2025SmarterTC}. Beyond code-level properties, \textit{Efficiency}-oriented rubric items capture system-level quality concerns such as computational time and space usage \citep{Jiang2024FromET, Rosenberg2002SoftwareQM, Curtis2022MeasuringTS}.

\paragraph{A subset of rubric items extend beyond traditional code quality methods.}
\label{app:extends-se-frameworks}
\textit{Explanatory/Ethical Awareness} extends existing notions of documentation \citep{Messer2023AutomatedGA, Messer2024HowCA, Menolli2025EducationalIF, Rai2022ARO}, by introducing ethical considerations, such as privacy, fairness, and societal impact, concerns largely missing from technically focused rubrics. \textit{User-Centeredness} has limited precedent; although usability appears in ISO/IEC 9126 \citep{Bishop2024EvaluatingSC} and productivity-oriented metrics exist \citep{Hariharan2025SemanticME}, our rubric extends beyond usability and efficiency to emphasize empathetic human–computer interaction and focus on problem context. \textit{Creativity/Innovation} represents the strongest departure from traditional frameworks, which prioritize adherence to established patterns (another rubric item \textit{Standards/Conventions}) and correct use of language idioms \citep{Ernst2017WhatTF} over novelty. The broader literature rarely treats creativity as a code quality criterion, reflecting a historical emphasis on predictability and maintainability despite creativity’s importance in domains such as optimization and novel algorithm design. \textit{Instruction Following} and \textit{Domain-Specific Detail} further reflect recent evaluation dimensions emerging from interactive, goal-conditioned code generation and the growing need for specialized knowledge in LLM applications across diverse domains.

\definecolor{humancolor}{RGB}{150,60,60}
\definecolor{llmcolor}{RGB}{0,82,155}
\definecolor{hybridcolor}{RGB}{120,120,120}
\begin{table*}[t]
\centering
\scriptsize
\resizebox{\columnwidth}{!}{%
\begin{tabular}{p{4.5cm}p{5cm}p{5cm}}
\toprule
\textbf{Rubric Axis} & \textbf{Upper End of Axis} & \textbf{Lower End of Axis}
 \\
\midrule

\textbf{\textcolor{llmcolor}{Error and Context Management}} &
\parbox[t]{5cm}{Comprehensive error handling, fallback mechanisms.} &
\parbox[t]{5cm}{Minimal or no error handling.} \\
\addlinespace
&
\parbox[t]{5cm}{\ttfamily
if not file.exists(path):\\
\hspace*{1.5em}raise FileNotFoundError(path)
} &
\parbox[t]{5cm}{\ttfamily open(path)} \\
\addlinespace[0.6em]
\midrule

\textbf{\textcolor{llmcolor}{Completeness and Integrity}} &
\parbox[t]{5cm}{Ensures all essential components and functional integrity.} &
\parbox[t]{5cm}{Misses important parts, leading to gaps.} \\
\addlinespace
&
\parbox[t]{5cm}{\ttfamily
connect()\\
query()\\
close()
} &
\parbox[t]{5cm}{\ttfamily query()} \\
\addlinespace[0.6em]
\midrule

\textbf{\textcolor{llmcolor}{Explanatory and Ethical Awareness}} &
\parbox[t]{5cm}{Provides depth and considers ethical implications.} &
\parbox[t]{5cm}{Minimal explanation with no ethical consideration.} \\
\addlinespace
&
\parbox[t]{5cm}{\ttfamily
\# Mask PII before logging
} &
\parbox[t]{5cm}{\ttfamily print(user\_ssn)} \\
\addlinespace[0.6em]
\midrule

\textbf{\textcolor{llmcolor}{Engagement and User Interaction}} &
\parbox[t]{5cm}{Engages empathetically and responds to user context.} &
\parbox[t]{5cm}{Lacks engagement and ignores user perspective.} \\
\addlinespace
&
\parbox[t]{5cm}{\ttfamily "JSON or CSV?"} &
\parbox[t]{5cm}{\ttfamily "Done."} \\
\addlinespace[0.6em]
\midrule

\textbf{\textcolor{llmcolor}{Creativity and Innovation}} &
\parbox[t]{5cm}{Introduces novel problem decompositions or reframes the task in an original way.} &
\parbox[t]{5cm}{Applies standard patterns without rethinking the structure of the problem.} \\
\addlinespace
&
\parbox[t]{5cm}{\ttfamily
def solve(items):\\
\hspace*{1.5em}return groupby(normalize(items))
} &
\parbox[t]{5cm}{\ttfamily
result = []\\
for x in items:\\
\hspace*{1.5em}result.append(x)
} \\
\addlinespace[0.6em]
\midrule

\textbf{\textcolor{humancolor}{Conciseness and Simplicity}} &
\parbox[t]{5cm}{Minimal, straightforward solutions avoiding unnecessary complexity.} &
\parbox[t]{5cm}{Unnecessarily complex and verbose.} \\
\addlinespace
&
\parbox[t]{5cm}{\ttfamily return sum(xs)} &
\parbox[t]{5cm}{\ttfamily
total = 0\\
for i in xs:\\
\hspace*{1.5em}total += i\\
return total
} \\
\addlinespace[0.6em]
\midrule

\textbf{\textcolor{humancolor}{Flexibility and Generality}} &
\parbox[t]{5cm}{Adaptable, modular solutions that handle diverse inputs.} &
\parbox[t]{5cm}{Rigid, specific implementations without generality.} \\
\addlinespace
&
\parbox[t]{5cm}{\ttfamily
def load(path, fmt):\\
\hspace*{1.5em}\dots
} &
\parbox[t]{5cm}{\ttfamily
def load\_csv(path):\\
\hspace*{1.5em}\dots
} \\
\addlinespace[0.6em]
\midrule

\textbf{\textcolor{hybridcolor}{Syntax and Structural Consistency}} &
\parbox[t]{5cm}{Adheres to syntax rules with consistent structure.} &
\parbox[t]{5cm}{Contains syntax errors and inconsistent elements.} \\
\addlinespace
&
\parbox[t]{5cm}{\ttfamily
def add(a,b):\\
\hspace*{1.5em}return a+b
} &
\parbox[t]{5cm}{\ttfamily
def add(a b)\\
\hspace*{1.5em}return a+b
} \\
\addlinespace[0.6em]
\midrule

\textbf{\textcolor{hybridcolor}{Functional and Logical Alignment}} &
\parbox[t]{5cm}{Matches expected behavior / logic.} &
\parbox[t]{5cm}{Deviates from intended behavior / logic.} \\
\addlinespace
&
\parbox[t]{5cm}{\ttfamily
if x > 0:\\
\hspace*{1.5em}handle(x)
} &
\parbox[t]{5cm}{\ttfamily
if x < 0:\\
\hspace*{1.5em}handle(x)
} \\
\addlinespace[0.6em]
\midrule

\textbf{\textcolor{hybridcolor}{Explicitness and Clarity}} &
\parbox[t]{5cm}{Clear, self-explanatory approaches that minimize ambiguity.} &
\parbox[t]{5cm}{Obscure and requires deeper analysis to understand.} \\
\addlinespace
&
\parbox[t]{5cm}{\ttfamily user\_count = len(users)} &
\parbox[t]{5cm}{\ttfamily uc = len(u)} \\
\addlinespace[0.6em]
\bottomrule
\end{tabular}
}
\caption{\textbf{Rubric items produced by our pipeline for code completion.} Color indicates whether the rubric is \textcolor{llmcolor}{LLM-generated}, \textcolor{humancolor}{human-annotated}, or \textcolor{hybridcolor}{hybrid} (due to aggregation).}
\label{tab:copilot-rubric-items}
\end{table*}

\begin{table*}[t]
\centering
\scriptsize
\resizebox{\columnwidth}{!}{%
\begin{tabular}{p{4.5cm}p{5cm}p{5cm}}
\toprule
\textbf{Rubric Axis} & \textbf{Upper End of Axis} & \textbf{Lower End of Axis} \\
\midrule

\textbf{\textcolor{llmcolor}{User Interaction and Feedback Responsiveness}} &
\parbox[t]{5cm}{Adapts based on prior feedback.} &
\parbox[t]{5cm}{Ignores feedback and repeats defaults.} \\
\addlinespace
&
\parbox[t]{5cm}{\ttfamily
use\_json = False\\
emit\_yaml(data)
} &
\parbox[t]{5cm}{\ttfamily
emit\_json(data)
} \\
\addlinespace[0.6em]
\midrule

\textbf{\textcolor{llmcolor}{Modularity and Code Structure}} &
\parbox[t]{5cm}{Promotes modular, organized code.} &
\parbox[t]{5cm}{Integrated and disorganized.} \\
\addlinespace
&
\parbox[t]{5cm}{\ttfamily
def load(): \\
\hspace*{1.5em}\dots\\
def save(): \\
\hspace*{1.5em}\dots\\
} &
\parbox[t]{5cm}{\ttfamily
def run():\\
\hspace*{1.5em}load()\\
\hspace*{1.5em}save()
} \\
\addlinespace[0.6em]
\midrule

\textbf{\textcolor{llmcolor}{Domain-Specific Detail and Technical Creativity}} &
\parbox[t]{5cm}{In-depth, creative domain-aware solutions.} &
\parbox[t]{5cm}{Generic and conventional.} \\
\addlinespace
&
\parbox[t]{5cm}{\ttfamily
use\_btree\_index(keys)
} &
\parbox[t]{5cm}{\ttfamily
store\_list(keys)
} \\
\addlinespace[0.6em]
\midrule

\textbf{\textcolor{humancolor}{Code Explanation and Clarity}} &
\parbox[t]{5cm}{Provides clear, detailed explanations of code.} &
\parbox[t]{5cm}{Lacks clarity and detail in explanation.} \\
\addlinespace
&
\parbox[t]{5cm}{\ttfamily
\# Validate before write\\
if not ok(x): \\
\hspace*{1.5em}raise Err()
} &
\parbox[t]{5cm}{\ttfamily
do\_thing(x)
} \\
\addlinespace[0.6em]
\midrule

\textbf{\textcolor{humancolor}{Language and Terminology Appropriateness}} &
\parbox[t]{5cm}{Uses preferred language and terminology.} &
\parbox[t]{5cm}{Uses undesired or unexpected language.} \\
\addlinespace
&
\parbox[t]{5cm}{\ttfamily
def enqueue(job): \\
\hspace*{1.5em}\dots
} &
\parbox[t]{5cm}{\ttfamily
def push\_stuff(x):\\
\hspace*{1.5em}\dots
} \\
\addlinespace[0.6em]
\midrule

\textbf{\textcolor{humancolor}{Efficiency and Simplicity}} &
\parbox[t]{5cm}{Efficient and straightforward design.} &
\parbox[t]{5cm}{Resource-intensive and complex.} \\
\addlinespace
&
\parbox[t]{5cm}{\ttfamily return sum(xs)} &
\parbox[t]{5cm}{\ttfamily
total = 0\\
for i in range(len(xs)):\\
\hspace*{1.5em}total += xs[i]
} \\
\addlinespace[0.6em]
\midrule

\textbf{\textcolor{humancolor}{Focus and Conciseness}} &
\parbox[t]{5cm}{Emphasizes the requested change only.} &
\parbox[t]{5cm}{Includes irrelevant details.} \\
\addlinespace
&
\parbox[t]{5cm}{\ttfamily
\# Patch overflow\\
limit = min(n, MAX)
} &
\parbox[t]{5cm}{\ttfamily
\# Here is a full redesign\\
init() \\
connect() \\
} \\
\addlinespace[0.6em]
\midrule

\textbf{\textcolor{humancolor}{Error-Free and Clarity of Presentation}} &
\parbox[t]{5cm}{Clear, well-formatted, error-free.} &
\parbox[t]{5cm}{Contains errors and unclear formatting.} \\
\addlinespace
&
\parbox[t]{5cm}{\ttfamily
if x == 0:\\
\hspace*{1.5em}return None
} &
\parbox[t]{5cm}{\ttfamily
if x = 0:\\
\hspace*{1.5em}return
} \\
\addlinespace[0.6em]
\midrule

\textbf{\textcolor{hybridcolor}{Intent Alignment and Instruction Adaptability}} &
\parbox[t]{5cm}{Adheres to goals and integrates complex instructions.} &
\parbox[t]{5cm}{Deviates from goals and struggles with instructions.} \\
\addlinespace
&
\parbox[t]{5cm}{\ttfamily
\# Only update auth logic\\
update\_auth(token)
} &
\parbox[t]{5cm}{\ttfamily
\# Refactor everything\\
rewrite\_system()
} \\
\addlinespace[0.6em]
\midrule

\textbf{\textcolor{hybridcolor}{Completeness and Precision}} &
\parbox[t]{5cm}{Thorough and precise.} &
\parbox[t]{5cm}{Broad and underspecified.} \\
\addlinespace
&
\parbox[t]{5cm}{\ttfamily
open()\\
read()\\
close()
} &
\parbox[t]{5cm}{\ttfamily
handle\_file()
} \\
\addlinespace[0.6em]

\bottomrule
\end{tabular}
}
\caption{\textbf{Rubric items produced by our pipeline for chat-based coding.} Color indicates whether the rubric is \textcolor{llmcolor}{LLM-generated}, \textcolor{humancolor}{human-annotated}, or a \textcolor{hybridcolor}{hybrid} (due to aggregation).}
\label{tab:lmarena-rubric-items}
\end{table*}

\begin{table*}[t]
\centering
\scriptsize
\resizebox{\columnwidth}{!}{%
\begin{tabular}{p{4.5cm}p{5cm}p{5cm}}
\toprule
\textbf{Rubric Axis} & \textbf{Upper End of Axis} & \textbf{Lower End of Axis} \\
\midrule

\textbf{\textcolor{llmcolor}{Instruction Fidelity}} &
\parbox[t]{5cm}{Strictly follows templates and instructions.} &
\parbox[t]{5cm}{Interprets instructions flexibly.} \\
\addlinespace
&
\parbox[t]{5cm}{\ttfamily
\# Format: name,age,date\\
print("\{name\},\{age\},\{date\}")
} &
\parbox[t]{5cm}{\ttfamily
\# Close enough\\
print(name, age)
} \\
\addlinespace[0.6em]
\midrule

\textbf{\textcolor{llmcolor}{Contextual and Creative Adaptability}} &
\parbox[t]{5cm}{Uses context cues to choose an appropriate approach.} &
\parbox[t]{5cm}{Uses a fixed approach regardless of context.} \\
\addlinespace
&
\parbox[t]{5cm}{\ttfamily
if len(records) > 1000000:\\
\hspace*{1.5em}process\_stream(records)\\
} &
\parbox[t]{5cm}{\ttfamily
process\_in\_memory(records)
} \\
\midrule

\textbf{\textcolor{humancolor}{Visual Presentation}} &
\parbox[t]{5cm}{Uses contrast and spacing to ensure readability.} &
\parbox[t]{5cm}{Places text on low-contrast backgrounds, harming legibility.} \\
\addlinespace
&
\parbox[t]{5cm}{\ttfamily
\# High contrast\\
plt.text(0.5,0.5,"Warning", \\
color="black",bbox=dict(\\
facecolor="yellow"))} &
\parbox[t]{5cm}{\ttfamily
\# Low contrast\\
plt.text(0.5,0.5,"Warning", \\
color="lightgray",bbox=dict( \\
facecolor="lightgray"))} \\
\addlinespace[0.6em]
\midrule

\textbf{\textcolor{humancolor}{Data and Type Management}} &
\parbox[t]{5cm}{Preserves data type integrity and handles errors gracefully.} &
\parbox[t]{5cm}{Simplifies data types and neglects detailed error handling.} \\
\addlinespace
&
\parbox[t]{5cm}{\ttfamily
x: int = int(s)\\
if x < 0: raise ValueError()
} &
\parbox[t]{5cm}{\ttfamily
x = s\\
return x
} \\
\addlinespace[0.6em]
\midrule

\textbf{\textcolor{hybridcolor}{Modularity and Abstraction}} &
\parbox[t]{5cm}{Uses modular, abstract components and reasoning.} &
\parbox[t]{5cm}{Prefers integrated, concrete implementations.} \\
\addlinespace
&
\parbox[t]{5cm}{\ttfamily
def parse(x): \\ 
\hspace*{1.5em}\dots\\
def validate(y): \\ 
\hspace*{1.5em}\dots
} &
\parbox[t]{5cm}{\ttfamily
def run(x):\\
\hspace*{1.5em}parse(x) \\
\hspace*{1.5em}validate(x)
} \\
\addlinespace[0.6em]
\midrule

\textbf{\textcolor{hybridcolor}{Conformance to Standards}} &
\parbox[t]{5cm}{Adheres to established standards and practices.} &
\parbox[t]{5cm}{Deviates from conventions with non-standard approaches.} \\
\addlinespace
&
\parbox[t]{5cm}{\ttfamily
class UserService: \\
\hspace*{1.5em}\dots
} &
\parbox[t]{5cm}{\ttfamily
class userservice123: \\
\hspace*{1.5em}\dots
} \\
\addlinespace[0.6em]
\midrule

\textbf{\textcolor{hybridcolor}{Correctness and Precision}} &
\parbox[t]{5cm}{Ensures logical and factual accuracy, focusing on details.} &
\parbox[t]{5cm}{Contains inaccuracies with broader strokes.} \\
\addlinespace
&
\parbox[t]{5cm}{\ttfamily
if n \% 2 == 0:\\
\hspace*{1.5em}even += 1
} &
\parbox[t]{5cm}{\ttfamily
if n > 0:\\
\hspace*{1.5em}even += 1
} \\
\addlinespace[0.6em]
\midrule

\textbf{\textcolor{hybridcolor}{Explicitness and Clarity}} &
\parbox[t]{5cm}{Provides clear, detailed documentation and explicit code elements.} &
\parbox[t]{5cm}{Lacks clarity with sparse documentation.} \\
\addlinespace
&
\parbox[t]{5cm}{\ttfamily
\# Count active users\\
active\_users = len(u)
} &
\parbox[t]{5cm}{\ttfamily
a = len(u)
} \\
\addlinespace[0.6em]
\midrule

\textbf{\textcolor{hybridcolor}{Brevity and Conciseness}} &
\parbox[t]{5cm}{Delivers clear, concise responses without redundancies.} &
\parbox[t]{5cm}{Includes verbose or superfluous content.} \\
\addlinespace
&
\parbox[t]{5cm}{\ttfamily return sum(xs)} &
\parbox[t]{5cm}{\ttfamily
total = 0\\
for i in xs:\\
\hspace*{1.5em}total += i\\
return total
} \\
\addlinespace[0.6em]
\midrule

\textbf{\textcolor{hybridcolor}{Robustness and Error Handling}} &
\parbox[t]{5cm}{Offers resilient solutions with comprehensive error management.} &
\parbox[t]{5cm}{Fragile solutions with basic error handling.} \\
\addlinespace
&
\parbox[t]{5cm}{\ttfamily
try: \\
\hspace*{1.5em}load(p)\\
except IOError: \\
\hspace*{1.5em}fallback()
} &
\parbox[t]{5cm}{\ttfamily load(p)} \\
\addlinespace[0.6em]

\bottomrule
\end{tabular}
}
\caption{\textbf{Rubric items produced by our pipeline for code edits.} Color indicates whether the rubric is \textcolor{llmcolor}{LLM-generated}, \textcolor{humancolor}{human-annotated}, or a \textcolor{hybridcolor}{hybrid} (due to aggregation).}
\label{tab:editbench-rubric-items}
\end{table*}

\begin{table*}[h]
\centering
\small
\setlength{\tabcolsep}{8pt}
\resizebox{\columnwidth}{!}{%
\begin{tabular}{
  >{\centering\arraybackslash}p{3.5cm}
  >{\centering\arraybackslash}p{2.8cm}
  >{\centering\arraybackslash}p{2.8cm}
  >{\centering\arraybackslash}p{2.8cm}
  >{\centering\arraybackslash}p{1.2cm}
}
\toprule
\textbf{Theme} & \textbf{Code Completion} & \textbf{Chat-based Coding} & \textbf{Instructed Code Edits} & \textbf{Scope} \\
\midrule
\textbf{Clarity / Explicitness} & Explicitness and Clarity  & Code Explanation and Clarity & Explicitness and Clarity & All \\
\midrule
\textbf{Conciseness} & Conciseness and Simplicity & Focus and Conciseness & Brevity and Conciseness & All \\
\midrule
\textbf{Correctness / Precision} & Functional and Logical Alignment & Completeness and Precision & Correctness and Precision & All \\
\midrule
\textbf{Modularity / Structure} & Flexibility and Generality  & Modularity and Code Structure & Modularity and Abstraction & All \\
\midrule
\textbf{Error Handling / Robustness} & Error and Context Management & Error-Free and Clarity of Presentation & Robustness and Error Handling & All \\
\midrule
\textbf{User-Centeredness} & Engagement and User Interaction & User Interaction and Feedback Responsiveness  & Contextual and Creative Adaptability & All \\
\midrule
\midrule
\textbf{Creativity / Innovation} & Creativity and Innovation & Domain-Specific Detail and Technical Creativity & -- & Two \\
\midrule
\textbf{Completeness} & Completeness and Integrity & Completeness and Precision & -- & Two \\
\midrule
\textbf{Instruction Following} & -- & Intent Alignment and Instruction Adaptability & Instruction Fidelity & Two \\
\midrule
\textbf{Standards / Conventions} & -- & Language and Terminology Appropriateness & Conformance to Standards & Two \\
\midrule
\textbf{Efficiency} & Conciseness and Simplicity & Efficiency and Simplicity & -- & Two \\
\midrule
\textbf{Presentation / Formatting} & -- & Error-Free and Clarity of Presentation & Visual Presentation & Two \\
\midrule
\midrule
\textbf{Syntax / Structural Consistency} & Syntax and Structural Consistency & -- & -- & One \\
\midrule
\textbf{Explanatory / Ethical Awareness} & Explanatory and Ethical Awareness & -- & -- & One \\
\midrule
\textbf{Data / Type Management} & -- & -- & Data and Type Management & One \\
\midrule
\textbf{Domain-Specific Detail} & -- & Domain-Specific Detail & -- & One \\
\bottomrule
\end{tabular}
}
\caption{\textbf{Several core themes recur across datasets, but each modality also contributes unique axes.} Generated rubric items across code completion, chat, and code edits, highlighting evaluation criteria shared across all datasets, shared by two datasets, or unique to one dataset.}
\label{table:rubric-generations}
\end{table*}

\end{document}